\documentclass[10pt]{report}
\usepackage[CJKbookmarks]{hyperref}
\usepackage{amsmath}
\usepackage{bm}
\usepackage{amssymb}
\usepackage{bbm}
\usepackage{cite}
\usepackage{graphicx}
\usepackage{simplewick}
\usepackage{mathrsfs}
\usepackage{centernot}
\usepackage{fullpage}
\usepackage{graphicx}
\usepackage{amssymb}
\usepackage{fancyvrb}
\usepackage{listings}
\usepackage{xcolor}
\usepackage{float}
\usepackage{dsfont}
\usepackage{longtable}
\usepackage{multirow}
\usepackage{courier}
\usepackage[T1]{fontenc}
\usepackage{indentfirst}
\begin{document}

\title{Revisiting Empirical Bayes Methods and Applications to Special Types of Data}
\author{Xiuwen Duan ~~300144074\\
Master of Science, University of Ottawa}
\maketitle
\clearpage
\tableofcontents
\listoffigures
\listoftables
\pagenumbering{arabic}

\abstract
\noindent
Empirical Bayes methods have been around for a long time and have a wide range of applications. These methods provide a way in which historical data can be aggregated to provide estimates of the posterior mean. This thesis revisits some of the empirical Bayesian methods and develops new applications.  We first look at a linear empirical Bayes estimator and apply it on ranking and symbolic data. Next, we consider Tweedie's formula and show how it can be applied to analyze a microarray dataset. The application of the formula is simplified with the Pearson system of distributions. Saddlepoint approximations enable us to generalize several results in this direction. The results show that the proposed methods perform well in applications to real data sets.
\newpage
\section*{Acknowledgement}
\noindent
I would like to express my sincere gratitude to my supervisor Professor Mayer Alvo for his patient guidance and continuous support during the process of thesis writing. He steered me in the right direction and helped to solve several bottleneck problems. He taught me a lot during this journey, not only through his profound knowledge but also by his optimistic altitude. I enjoyed these last two years learning Mathematics and Statistics due to his inspiration. As well, I want to thank the previous Director of the graduate program, Professor Benoit Dionne, the present Director Professor Gilles Lamothe, for their patient and quick responses throughout the whole program. Then I would like to thank the department for providing a great academic environment. All the professors I met were kind and helpful. Finally yet importantly, I would like to thank my parents, my girl friend Yalin Liu and friends for all the encouragement and spiritual support along the way.

\newpage

\chapter{Introduction}

In its most basic description, the study of statistical inference begins by specifying the sampling distribution $g(x|\theta)$ of an observable random variable $x$ which may be a vector. Here $\theta$ is not observed.
The goal then is to either estimate $\theta$ or to test hypotheses about $\theta$ on the basis of a random sample drawn from $g(x|\theta).$
The Bayesian approach supposes that $\theta$, is not a fixed constant but rather a random variable as well having a density $\pi(\theta)$.
Inference in that case is based on the posterior distribution of $\theta$ conditional on the observed value of x, and in our notation is written simply as $g(\theta|x).$
In general, the joint probability density function can be expressed as a product of two densities that are referred to as the sampling distribution (or data distribution) $g(x|\theta)$ and the prior $\pi(\theta)$ respectively\cite{27}
\begin{align*}
g(\theta,x)=g(x|\theta)\pi(\theta)
\end{align*}

Simply conditioning on the known value of the data x, using the basic property of conditional probability known as Bayes Theorem, yields the posterior density:
\begin{align}
g(\theta|x)=\frac{g(\theta,x)}{g(x)}=\frac{g(x|\theta)\pi(\theta)}{g(x)}
\end{align}
where $g(x)=\int g(x|\theta)\pi(\theta)d\theta$ , which is the integral over all possible values of $\theta$.

In the pure Bayesian approach, we need to specify the prior distribution $\pi(\theta)$.
When the prior distribution is unspecified, empirical Bayes methods come into play.
The empirical Bayes (EB) approach can be thought a way of dealing with data arising in a sequence of similar (or same type of) experiments\cite{40}.
Inference about the current component $\left(x_{n},\theta_{n}\right)$
take into consideration all other components $\left(x_{1},\theta_{1}\right),...,\left(x_{n-1},\theta_{n-1}\right)$.
Therefore, empirical Bayes methods make the historical data also contribute to the current data estimates.\\

Empirical Bayes approaches have been around for a long time and have been shown to be powerful data analytic tools\cite{13}.
They have a wide range of applications in real life, including estimating baseball batting averages, assessing performances of competitors and comparing scores of different schools - to name a few.
Our objectives in this thesis are to revisit some of the empirical Bayes methods and to exploit certain techniques to enhance their application.

In Chapter 2, we consider the linear empirical Bayes estimator proposed by Robbins and generalize it to obtain a vector form. We will apply these results to ranking data and to symbolic data in real life data sets.

In Chapter 3, we introduce Tweedie's formula and its multivariate version.
This formula enables us to compute the posterior mean of $\theta$ given a current value of $x.$ The formula requires knowledge of the derivative of the logarithm of the marginal of $x$.
We outline a method based on the Pearson system of distributions to determine this derivative in terms of the first four moments of the data.
Then we will illustrate how the estimation proceeds with a microarray example.
We also consider a generalization based on the saddlepoint approximation.
These results are applied on ranking data and we obtain promising results.

In Chapters 4, we conclude by summarizing the findings of this thesis and suggest two directions for further research and investigation.

\chapter{Linear Empirical Bayes method}
\section{Linear Empirical Bayes Estimator}

Consider the situation where we begin by observing $x_{1}$ whose cumulative distribution, indexed by a parameter $\theta_{1}$, is $F\left(x_{1}|\theta_{1}\right).$
Our problem is to estimate $\theta_{1}.$ At some later time, we observe independently $x_{2}$ whose cumulative distribution, indexed by a parameter $\theta_{2}$, is $F\left(x_{2}|\theta_{2}\right).$
Once again we wish to estimate $\theta_{2}$. We continue and we observe $n$ random variables $x_{1},x_{2},...,x_{n}$ and at stage $n$ we wish to estimate $\theta_{n}.$
In the classical statistics framework, our best estimate for $\theta_{n}$ would be based entirely on $x_{n}$ alone. Intuitively, this would not be a very good procedure because it would be based on a sample of size one.
On the other hand, a classical Bayesian approach would at stage $n$ begin by imposing a known prior distribution $\pi\left(\theta\right)$ on $\theta$ (which would be the same for all stages) and set about to compute the posterior mean of $\theta$ given $x_{1},x_{2},...,x_{n}$
\begin{align}
\frac{\int\theta\pi(\theta)dF(x_{1},x_{2},...,x_{n}|\theta)}{\int\pi(\theta)dF(x_{1},x_{2},...,x_{n}|\theta)}
\end{align}

In the classical Bayes approach if we assume the $\theta_{i}$'s are all different, the posterior mean of $\theta_{i}$ take account only $x_{i}$.
Our example on gene expression levels in the next section will make this clear. There $\theta_{i}$ is specific to ith individual.
Empirical Bayes methods were introduced to deal with the situation where the prior distribution $\pi(\theta)$ in the Bayesian problem above was unspecified.
The empirical Bayes methods often provide superior estimates of parameters than either the classical model or the ordinary Bayes model\cite{13}. Their origins can be traced
back to work by von Mises in the 1940's and Herbert Robbins in 1956\cite{43}.
In this section, we explore Robbins's formulation which sometimes referred to as "non-parametric empirical Bayes".
We emphasize here that there is no attempt to estimate the prior.

In 1984, H. Robbins proposed a linear empirical Bayes estimator to estimate the posterior mean and variance\cite{46}.
His goal was to estimate the means and variances from populations whose structure is unknown or partly known.
Our contribution here is to generalize this method to its vector form and apply it to special types of data.

Let $(\varphi,x)$ be a random vector where the vector $x$ corresponds
to observed data and the vector $\varphi$ is the unknown parameter.
In our setting the parameter $\varphi$ stands for the conditional mean of $x$ (This is why we use the notation $\varphi$ instead of $\theta$), so we have the assumption:
\begin{align}
\varphi=E[x|\varphi].
\end{align}

We wish to estimate $\varphi$ by a function t of x for the purpose to minimize the mean square error.

Since
\begin{align}
E[(t-\varphi)(t-\varphi)^{T}]\geqslant E[Var(\varphi|x)],
\end{align}
with equality only if $P(t=E[\varphi|x])=1$. So $E[\varphi|x]$ is the best estimator for $\varphi$.

While the joint distribution of $(\varphi,x)$ is unknown, $E[\varphi|x]$ cannot be obtained explicitly.
We consider the estimator $t^{*}(x)$ by a linear vector of the form
\begin{align}
t^{*}(x)=a+Bx, \label{eq:Linear function}
\end{align}
where $a$ is a vector and $B$ is a matrix.

The reason why we choose this linear function is that it is the easiest way to work with in the problem of minimizing $E[(t-\varphi)(t-\varphi)^{T}]$.

Let
\begin{align}
\phi(a,B)=E[(t^{*}(x)-\varphi)(t^{*}(x)-\varphi)^{T}]=E[(a+Bx-\varphi)(a+Bx-\varphi)^{T}].  \label{eq:Minimizer function}
\end{align}

Re-writing, we have upon subtracting and adding $\left(BEx-E\varphi\right)$,
\begin{align}
a+Bx-\varphi=\left(a-E\varphi+BEx\right)+\left(E\varphi-\varphi-BEx+Bx\right).
\end{align}

Hence, equation (\ref{eq:Minimizer function}) becomes
\begin{align}
\phi\left(a,B\right)= & \left[\left(a-\left(E\varphi-BEx\right)\right)\left(a-\left(E\varphi-BEx\right)\right)^{T}\right] \nonumber \\
 & +E\left[\left(B\left(x-Ex\right)\right)\left(B\left(x-Ex\right)\right)^{T}\right] \nonumber \\
 & +E\left[\left(\varphi-E\varphi\right)\left(\varphi-E\varphi\right)^{T}\right] \nonumber \\
 & -E\left[\left(B\left(x-Ex\right)\right)\left(\varphi-E\varphi\right)^{T}\right] \nonumber \\
 & -E\left[\left(\varphi-E\varphi\right)\left(B\left(x-Ex\right)\right){}^{T}\right] \nonumber \\
= & \left(a-\left(E\varphi-BEx\right)\right)\left(a-\left(E\varphi-BEx\right)\right)^{T} \nonumber \\
 & +B\varSigma_{x}B^{T}-B\varSigma_{x\varphi}-\varSigma_{x\varphi}B^{T}+\varSigma_{\varphi}.
\end{align}
where $\varSigma_{x},\varSigma_{x\varphi}$ are the variance and covariance
matrices of $x$ and $\left(x,\varphi\right)$ respectively.

Then, we may write
\begin{align}
\phi\left(a,B\right) & =  \left(a-\left(E\varphi-BEx\right)\right)\left(a-\left(E\varphi-BEx\right)\right)^{T} \nonumber \\
 & +  \left(B\varSigma_{x}^{1/2}-\varSigma_{x\varphi}\varSigma_{x}^{-1/2}\right)\left(B\varSigma_{x}^{1/2}-\varSigma_{x\varphi}\varSigma_{x}^{-1/2}\right)^{T}
 \nonumber \\
 & +  \varSigma_{\varphi}-\left(\varSigma_{x}^{-1/2}\varSigma_{x\varphi}\right)\left(\varSigma_{x}^{-1/2}\varSigma_{x\varphi}\right)^{T}.
\end{align}

It follows that $\phi(a,B)$ is minimized with respect
to $a$ and $B$ whenever
\begin{align}
a & = ~\left(E\varphi-BEx\right), \nonumber \\
B & = ~\varSigma_{x\varphi}\varSigma_{x}^{-1}. \label{eq:a&B solution}
\end{align}

The minimum of equation (\ref{eq:Minimizer function}) over $(a,B)$ is then
\begin{align}
min\:\phi\left(a,B\right)=\varSigma_{\varphi}-\left(\varSigma_{x}^{-1/2}\varSigma_{x\varphi}\right)\left(\varSigma_{x}^{-1/2}\varSigma_{x\varphi}\right)^{T}.
\end{align}

Applying the law of total expectation to $\varphi=E[x|\varphi]$, we get:
\begin{align}
Ex=E\varphi;~\varSigma_{x\varphi}=\varSigma_{\varphi}.
\end{align}

Plug equation (\ref{eq:a&B solution}) in equation (\ref{eq:Linear function}), the estimator $t^{*}(x)$ for $\varphi$ is given by
\begin{align}
t^{*}(x) & = E\varphi+\varSigma_{x\varphi}\varSigma_{x}^{-1}\left(x-Ex\right) \nonumber \\
 & = E\varphi+\varSigma_{\varphi}\varSigma_{x}^{-1}\left(x-Ex\right). \label{eq:Linear estimator}
\end{align}

$E\varphi$ and $Ex$ can be estimated by the sample mean.\\

Next, we consider how to use the data to estimate $\Sigma_{\varphi}$
and $\Sigma_{x}$ which are the variance-covariance matrices of x
and $\varphi$ respectively.

Let $F_{1},...,F_{N}$ be i.i.d. random
distribution functions and let $\tilde{X_{i}}=[X_{i1},X_{i2},...,X_{in_{i}}]$
be a random sample with size $n_{i}$ from distribution $F_{i}$.
Here, in our application the $X_{ij}$ for $i=1,...,N$ and $j=1,...,n_{i}$ are random
vectors. For each distribution $F_{i}$, define the random parameters:
\begin{align}
\varphi_{i} & =E[X_{ij}|F_{i}], \nonumber \\
\sigma_{i}^{2} & =Var[X_{ij}|F_{i}].
\end{align}

Then define the following statistics:
\begin{align}
 & \bar{X_{i}}=\frac{1}{n_{i}}\overset{n_{i}}{\underset{j}{\sum}}X_{ij},
 ~~~~~~~~~~~~~~~~~~~~~~~~~~~~~~~~~\bar{X}=\frac{1}{N}\overset{N}{\underset{i}{\sum}}\bar{X_{i}}, \nonumber \\
 & S_{i}^{2}=\frac{1}{n_{i}-1}\overset{n_{i}}{\underset{j}{\sum}}(X_{ij}-\bar{X_{i}})(X_{ij}-\bar{X_{i}})^{T},
 ~~ S^{2}=\frac{1}{N}\overset{N}{\underset{i}{\sum}}S_{i}^{2}, \nonumber \\
 & U^{2}=\frac{1}{N-1}\overset{N}{\underset{i}{\sum}}(\bar{X_{i}}-\bar{X})(\bar{X_{i}}-\bar{X})^{T},
 ~~~~~~ v=\frac{1}{N}\overset{N}{\underset{i}{\sum}}\frac{1}{n_{i}}.
\end{align}

Note that
\begin{align}
E[\bar{X_{i}}] & =E[\varphi_{i}]=E[\bar{X}], \label{eq:Expectations of sample mean}\\
E[S_{i}^{2}] & =E[\sigma_{i}^{2}]=E[S^{2}], \label{eq:Expectations of sample variance}\\
\Sigma_{X_{i}} & =\frac{ES_{i}^{2}}{n_{i}}+\Sigma_{\varphi_{i}}. \label{eq:Relationship between sample mean and variance}
\end{align}

Equations (\ref{eq:Expectations of sample mean}) and (\ref{eq:Expectations of sample variance}) come from properties
of the parameters whereas (\ref{eq:Relationship between sample mean and variance}) is the result of the Law of total variance.

Further, we have:
\begin{align*}
(N-1)U^{2}= & \overset{N}{\underset{i=1}{\sum}}(\bar{X_{i}}-\bar{X})(\bar{X_{i}}-\bar{X})^{T}\\
= & \overset{N}{\underset{i=1}{\sum}}(\bar{X_{i}}-\bar{X})(\bar{X_{i}}^{T}-\bar{X}^{T})\\
= & \overset{N}{\underset{i=1}{\sum}}(\bar{X_{i}}\bar{X_{i}}^{T})-\overset{N}{\underset{i=1}{\sum}}(\bar{X_{i}}\bar{X}^{T}+\bar{X}\bar{X_{i}}^{T})+N\bar{X}\bar{X}^{T}\\
= & {\underset{i}{\sum}}(\bar{X_{i}}\bar{X_{i}}^{T})-N\bar{X}\bar{X}^{T}.
\end{align*}

Taking expectation:
\begin{align*}
(N-1)EU^{2}= & {\underset{i}{\sum}}E(\bar{X_{i}}\bar{X_{i}}^{T})-EN(\bar{X}\bar{X}^{T})\\
= & {\underset{i}{\sum}}((\Sigma_{X_{i}})+E\bar{X_{i}}(E\bar{X_{i}})^{T})-N\Big\{\frac{1}{N^{2}}{\underset{i}{\sum}}(\Sigma_{X_{i}})+E\bar{X}(E\bar{X})^{T}\Big\}\\
= &
\frac{N-1}{N}{\underset{i}{\sum}}(\Sigma_{X_{i}}) +{\underset{i}{\sum}}E\bar{X_{i}}(E\bar{X_{i}})^{T}-N\big\{E\bar{X}(E\bar{X})^{T}\big\},\\
EU^{2}= & \frac{1}{N}{\underset{i}{\sum}}(\Sigma_{X_{i}})\\
= & vES^{2}+\Sigma_{\varphi_{i}}.
\end{align*}

Thus we obtain
\begin{align}
\Sigma_{\varphi_{i}} & =E(U^{2}-vS^{2}), \nonumber \\
\Sigma_{X_{i}} & =E(U^{2}-vS^{2})+\frac{1}{n_{i}}ES^{2}. \label{eq:Estimation of variance and covariance matrices}
\end{align}

Substitute equation (\ref{eq:Estimation of variance and covariance matrices}) into equation (\ref{eq:Linear estimator}), the linear empirical Bayes estimator $\hat{t_{i}}$ is then given by
\begin{align}
\hat{t_{i}}= ~ EX_{i}+E(U^{2}-vS^{2})[E(U^{2}-vS^{2})+\frac{1}{n_{i}}ES_{i}^{2}]^{-1}(X_{i}-EX_{i}).
\end{align}

Dropping all the expectations, we can approximate $\hat{t_{i}}$, for large
$N$ by
\begin{align}
\hat{t_{i}}= ~ \bar{X}+(U^{2}-vS^{2})^{+}[(U^{2}-vS^{2})^{+}+\frac{1}{n_{i}}S^{2}]^{-1}(\bar{X_{i}}-\bar{X}), \label{eq:Linear EB estimates vector form}
\end{align}
where $(U^{2}-vS^{2})^{+}$ means the diagonal values are $max\{0,diag(U^{2}-vS^{2})\}.$
This is used instead of $U^{2}-vS^{2}$ because the variance being estimated should be non-negative. The $\hat{t_{i}}$ represent our linear empirical Bayes estimates for the mean $\varphi_{i}$ of the vectors $X_{ij}$ in group i.

\section{Linear EB Application on Ranking data}

Ranking data often arise when it is desired to rank a set of individuals or objects according to some criterion.\cite{1}
A ranking establishes a relationship among a set of items such that, for any pairs of items, the first is either "ranked higher than", "ranked lower than" or "ranked equal to" the second. If two items are the same in rank it is called a tie.
Ranking data are usually encountered in practice when individuals are asked to rank a set of $t$ items, which may be competitors in a game, types of food, preferences of movies, etc.
Such data may be observed directly or come from the transformation of continuous or discrete data in a nonparametric analysis.\cite{1}

In statistics, ranking is the data which replace their original numerical values with their ranks after sorting.
Rankings make it possible to evaluate complex information using nonparametric statistical methods which reduces the detailed procedures analysing the original values.
This study aims to apply empirical Bayes methods to ranking data, both parametric and nonparametric. \\

\begin{description}
\item [{Application}] Canadian University Ranking data
\end{description}
\addcontentsline{toc}{section}{Application: Canadian University Ranking data}

As an example, we consider the data reported in McLean's Magazine
2018-2020 on the Canadian Medical University rankings.\footnote{\emph{https://www.macleans.ca/hub/education-rankings/}}

\begin{table}[H]
\centering \resizebox{\textwidth}{!}{ %
\begin{tabular}{|c|c|c|c|c|c|c|c|c|}
\hline
 & McGill  & Toronto  & UBC  & McMaster  & Alberta  & Queen  & Western  & Dalhousie\tabularnewline
\hline
Overall  & 1  & 2  & 3  & 4  & 5  & 5  & 7  & 8\tabularnewline
Student Awards  & 1  & 3  & 2  & 8  & 6  & 5  & 9  & 4\tabularnewline
Student/Faculty Ratio  & 5  & 13  & 4  & 14  & 7  & 15  & 10  & 2\tabularnewline
Faculty Awards  & 3  & 1  & 5  & 7  & 4  & 2  & 9  & 6\tabularnewline
Humanities Grants  & 1  & 3  & 2  & 6  & 7  & 8  & 11  & 4\tabularnewline
Medical/Science Grants  & 3  & 1  & 5  & 4  & 6  & 6  & 8  & 13\tabularnewline
Citations  & 3  & 1  & 2  & 4  & 6  & 13  & 9  & 10\tabularnewline
Total Research Dollars  & 3  & 1  & 7  & 2  & 4  & 10  & 11  & 14\tabularnewline
Operating Budget  & 10  & 6  & 4  & 9  & 1  & 8  & 12  & 5\tabularnewline
Library Expenses  & 2  & 4  & 14  & 13  & 5  & 5  & 8  & 10\tabularnewline
Library Acquisitions  & 1  & 10  & 14  & 3  & 11  & 12  & 7  & 8\tabularnewline
Scholarships  & 1  & 7  & 11  & 14  & 9  & 5  & 3  & 3\tabularnewline
Student Services  & 11  & 3  & 12  & 1  & 15  & 8  & 6  & 9\tabularnewline
\hline
\hline
 & Calgary  & Montreal  & Ottawa  & Laval  & Sherbrooke  & Manitoba  & Saskatchewan  & \tabularnewline
\hline
Overall  & 9  & 10  & 11  & 12  & 13  & 14  & 15  & \tabularnewline
Student Awards  & 9  & 11  & 7  & 12  & 13  & 14  & 15  & \tabularnewline
Student/Faculty Ratio  & 3  & 11  & 12  & 9  & 1  & 8  & 6  & \tabularnewline
Faculty Awards  & 12  & 10  & 8  & 11  & 15  & 12  & 14  & \tabularnewline
Humanities Grants  & 13  & 5  & 10  & 9  & 14  & 12  & 15  & \tabularnewline
Medical/Science Grants  & 11  & 9  & 2  & 10  & 15  & 14  & 12  & \tabularnewline
Citations  & 8  & 5  & 7  & 12  & 15  & 11  & 14  & \tabularnewline
Total Research Dollars  & 8  & 5  & 9  & 6  & 15  & 12  & 13  & \tabularnewline
Operating Budget  & 3  & 13  & 11  & 15  & 14  & 7  & 2  & \tabularnewline
Library Expenses  & 10  & 8  & 10  & 5  & 14  & 1  & 2  & \tabularnewline
Library Acquisitions  & 6  & 15  & 4  & 5  & 2  & 13  & 9  & \tabularnewline
Scholarships  & 6  & 10  & 2  & 11  & 15  & 8  & 13  & \tabularnewline
Student Services  & 7  & 13  & 4  & 10  & 5  & 2  & 14  & \tabularnewline
\hline
\end{tabular}} \caption{Maclean's 2020 Canadian Medical Universities Ranking}
\label{tb:Maclean's 2020 Canadian Medical Universities Ranking}
\end{table}

We analyze the ranking data for 12 different criteria of Canadian top 15 medical universities from 2018 to 2020.
As the ranking of one university in different years should have similar pattern, we view different years as different distributions. And the criteria are viewed as samples of a distribution, as they all contribute to the university's overall ranking.
We estimate the underlying parameter $\varphi$ the university has over the years, and rank it to predict the overall ranking of Canadian Medical Universities in 2020. The rankings are given in vector form, so we can directly apply our estimator.
The EB estimator gives us the estimated posterior means of $\varphi_{\textrm{2020}}$, which we then we rank. After the computation, the 2020 university's ranking is given by:

\begin{table}[H]
\centering \resizebox{\textwidth}{!}{ %
\begin{tabular}{|c|c|c|c|c|c|c|c|c|c|c|c|c|c|c|c|}
\hline
 & McGill  & Toronto  & UBC  & McMaster  & Alberta  & Queen  & Western  & Dalhousie  & Calgary  & Montreal  & Ottawa  & Laval  & Sherbrooke  & Manitoba  & Saskatchewan\tabularnewline
\hline
Maclean's  & 1  & 2  & 3  & 4  & 5  & 6  & 7  & 8  & 9  & 10  & 11  & 12  & 13  & 14  & 15\tabularnewline
\hline
Our EB estimates  & 1  & 2  & 6  & 3  & 5  & 7  & 10  & 8  & 9  & 11  & 4  & 12  & 15  & 13  & 14\tabularnewline
\hline
\end{tabular}} \caption{Comparison of Maclean's 2020 university ranking with our empirical bayes
estimates}
\label{tb:Comparison of Maclean's 2020 university ranking with our empirical bayes estimates}
\end{table}

As can be seen from Table \ref{tb:Comparison of Maclean's 2020 university ranking with our empirical bayes estimates},
our result appears very similar to the true Maclean's Magazine overall ranking.
McGill University and the University of Toronto are still ranked 1 and 2 respectively.
Western University, Dalhousie University, the University of Calgary and the University of Montreal still occupy rank 7-10, while Laval Univerisity, the University of Sherbrooke, the University of Manitoba and the University of Saskatchewan still rank 12-15.
In our formulation University of Ottawa is ranked 4 as opposed to 11 in Maclean's Magazine, which means it is underestimated by the Maclean's. If we consider all the criteria together equally and combine the data over the years, we think the University of Ottawa will be given
a more favorable ranking.

\section{Linear EB Application on Symbolic data}

Data nowadays is available in businesses, the internet, government
databases and is quite voluminous, challenging the capacity of computers.\cite{22}
The information age provides us more and more opportunities to deal with large data sets.
It is therefore important to summarize these data in ways that emphasize the underlying relationships and extract new knowledge from them.\cite{21}
We aim to analyze, visualize, classify and reduce the information from a more complicated type of data which we label symbolic data as they are structured and contain internal variation.\cite{21}

In the symbolic data paradigm, data is usually obtained from three different sources: multi-valued variables, interval-valued variables and modal variables.
It may take the form of classes which can be described by intervals, histograms, sets of categories etc.\cite{9}
For example, rather than specifying a single number for systolic blood pressure, one may specify an interval which can incorporate the uncertainty about the measurements.
As well symbolic data for a group of objects are summarized by an interval indicating the range of values.\cite{8}

The methods used to analyze symbolic data are extensions of the classical statistical methods which are adapted
to deal with symbolic data sets with special types of variables.\cite{9}
Hence, some common inferential methods such as regression analysis, principal component analysis and clustering, can still be used on symbolic data.

We have previously discussed the application of linear EB to ranking data. In this section, we investigate the application of this method on a particular type of symbolic data, namely interval data.
\subsection{Introduction of Interval data}

The representation of data by means of intervals of values is becoming increasingly more frequent in different fields of application.
Intervals appear as a way to describe the uncertainty affecting the observed values. The uncertainty can be considered as the inability to obtain true values due to our limited knowledge of the model that fits the question.\cite{33}
%

Interval data may occur in many different situations. We may have 'native' interval data, describing ranges of variable values, for instance, income of individuals in a company, horses weight and size level in a competition; or imprecise data, coming from repeated measures.
Interval-valued data may also arise from aggregation of a large dataset into one of more manageable size, here ordinary values are transformed into intervals.\cite{8}
The nature of the aggregation would vary depending on the different research objectives.\cite{12}

Let's take an example to see how the aggregation works. Consider measuring blood pressure for a number
of retired people. Aggregating these over age produced intervals with each group containing differing data. Thus we can get the blood pressure information for different age groups.\\

\begin{table}[H]
\centering %
\begin{tabular}{|cc|c|cc|}
\hline
age  & systolic pressure  &  & age group  & systolic interval\tabularnewline
\hline
70  & 130  &  & 55-59  & (110,130)\tabularnewline
67  & 115  &  & 60-64  & (128.140)\tabularnewline
57  & 124  &  & 65-69  & (115,120)\tabularnewline
64  & 128  &  & 70-74  & (120,130)\tabularnewline
74  & 120  & $\Rightarrow$  &   & \tabularnewline
65  & 120  &  &  & \tabularnewline
56  & 130  &  &        &          \tabularnewline
59  & 110  &  &        &          \tabularnewline
60  & 140  &  &        &          \tabularnewline
\hline
\multicolumn{2}{|c}{Before}   &  &  \multicolumn{2}{c|}{Aggregated}\tabularnewline
\hline
\end{tabular}\caption{Demonstration of aggregation into interval data}
\label{tb:Demonstration of aggregation into interval data}
\end{table}

\subsubsection{Descriptive Statistics}

We generally describe the interval by its lower and upper bounds,
denoted l and u, or we can use the center and half-width, denoted c and r.
This indicates the unique property of interval data, which contains
a kind of uncertainty.

In order to analyze interval data in our context, we need to define some descriptive statistics.
We give a recap of the approach proposed by Bertrand and Goupil (2000)\cite{7}.
Suppose X is an interval-valued variable, it is
measured for each element of the basic set E=$\{1, . . . ,n\}$.
For each $k\in E$, we denote the interval X(w) by $[l_{w},u_{w}]$.

The empirical density function of a univariate interval-valued variable X is defined to be
\begin{align}
f(\xi)=\frac{1}{n}{\underset{w\in E}{\sum}}\frac{\mathbbm{1}_{w}(\xi)}{\|X(w)\|},~~ \xi\in\mathbb{R},
\end{align}
where $\mathbbm{1}_{w}(\cdot)$ is the indicator function and $\|X(w)\|$ is the length of that interval.

Thus we can compute the symbolic sample mean as:
\begin{align}
\bar{X} & =\int_{-\infty}^{+\infty}\xi f(\xi)d\xi \nonumber \\
 & =\frac{1}{n}{\underset{w\in E}{\sum}}\int_{-\infty}^{+\infty}\frac{\mathbbm{1}_{w}(\xi)}{||X(w)||}\xi d\xi \nonumber \\
 & =\frac{1}{n}{\underset{w\in E}{\sum}}\frac{1}{u_{w}-l_{w}}\int_{\xi\in X(w)}\xi d\xi \nonumber \\
 & =\frac{1}{2n}{\underset{w\in E}{\sum}}\frac{u_{w}^{2}-l_{w}^{2}}{u_{w}-l_{w}} \nonumber \\
 & =\frac{1}{n}{\underset{w\in E}{\sum}}\frac{l_{w}+u_{w}}{2}. \label{eq:Symbolic mean}
\end{align}

Then the symbolic sample variance can be obtained in similar way:
\begin{align}
S^{2} &=\int_{-\infty}^{+\infty}(\xi-\bar{X})^{2}f(\xi)d\xi \nonumber \\
 & =\int_{-\infty}^{+\infty}\xi^{2}f(\xi)d\xi-2\bar{X}\int_{-\infty}^{+\infty}\xi f(\xi)d\xi+\bar{X}^{2}\int_{-\infty}^{+\infty}f(\xi)d\xi \nonumber \\
 & =\frac{1}{n}{\underset{w\in E}{\sum}}\frac{1}{u_{w}-l_{w}}\cdot\frac{u_{w}^{3}-l_{w}^{3}}{3}-2\bar{X}\frac{1}{n}{\underset{w\in E}{\sum}}\frac{l_{w}+u_{w}}{2}+\bar{X}^{2} \nonumber \\
 & =\frac{1}{3n}{\underset{w\in E}{\sum}}(u_{w}^{2}+u_{w}l_{w}+l_{w}^{2})-\frac{1}{4n^{2}}{\underset{w\in E}{\sum}}(l_{w}+u_{w})^{2}. \label{eq:Symbolic variance}
\end{align}

We also need to define the symbolic sample covariance\cite{10}:
\begin{align}
Cov(X_{i},X_{j}) & =\int_{-\infty}^{\infty}(\xi_{i}-\bar{X_{i}})(\xi_{j}-\bar{X_{j}})f(\xi_{i},\xi_{j})d\xi_{i}d\xi_{j} \nonumber \\
 & =\frac{1}{n}{\underset{w\in E}{\sum}}\frac{1}{(u_{iw}-l_{iw})(u_{jw}-l_{jw})}\int\int_{(\xi_{i},\xi_{j})\in X(w)}\xi_{i}\xi_{j}d\xi_{i}d\xi_{j}-\bar{X_{i}}\bar{X_{j}} \nonumber \\
 & =\frac{1}{4n}{\underset{w\in E}{\sum}}(l_{iw}+u_{iw})(l_{jw}+u_{jw})-\frac{1}{4n^{2}}{\underset{w\in E}{\sum}}(l_{iw}+u_{iw}){\underset{w\in E}{\sum}}(l_{jw}+u_{jw}). \label{eq:Symbolic covariance}
\end{align}

With these computations in place, we continue with our contributions to proceed the application of the linear EB method to interval data.

\subsection{Linear EB estimator of interval data}

First we consider scalar form, which means our data are interval-valued variables, here each sample is a pair of interval bounds.

Suppose $\varphi$ now is the symbolic mean of an interval-valued variable, we want to estimate $\varphi$
by a linear function $t^{*}(x)=a+bx$, where a and b are a real numbers.
Let
\begin{align}
\phi(a,b) & =E(a+bx-\varphi)^{2} \nonumber \\
 & =(a-(E\varphi-bEx))^{2} + Var(x)(b-\frac{Cov(x,\varphi)}{Var(x)})^{2} \nonumber \\
 &~~+Var(\varphi)-\frac{Cov(x,\varphi)^{2}}{Var(x)}. \label{eq:Minimizer function linear}
\end{align}

Minimizing this function (\ref{eq:Minimizer function linear}), we obtain
\begin{align}
a=E\varphi-bEx, \nonumber \\
b=\frac{Cov(x,\varphi)}{Var(x)}. \label{eq:a&B solution interval}
\end{align}

Note that $E(x|\varphi)=\varphi$, so $E\varphi=Ex$ and $E(\varphi x)=E\varphi^{2}$
which leads to $Cov(x,\varphi)=Var(\varphi)$, we have the estimator
\begin{align}
t^{*}(x)=Ex+\frac{Var(\varphi)}{Var(x)}(x-Ex). \label{eq:Linear EB estimator interval}
\end{align}

Suppose we have N distributions $F_{1},...,F_{N}$, each population contains samples $X_{ij}$ for $j=1,...,n_{i}$ with sample size $n_{i}$.

Ex are estimated from the symbolic mean of the intervals. The estimation of $Var~x$ and $Var~\varphi$ comes similar with equation (\ref{eq:Estimation of variance and covariance matrices}) but with new definition of the symbolic sample mean (\ref{eq:Symbolic mean}) and variance (\ref{eq:Symbolic variance}),
\begin{align}
&\bar{x}_{i} =\textrm{symbolic sample mean of group}~i, \nonumber \\
&s_{i}^{2} =\textrm{symbolic sample variance of group}~i, \nonumber \\
&\bar{x} =\frac{1}{N}\overset{N}{\underset{i=1}{\sum}}\bar{x}_{i}, ~~~~~~~~~~~~~~~~s^{2}  =\frac{1}{N}\overset{N}{\underset{i=1}{\sum}}s_{i}^{2}, \nonumber \\
&u^{2} =\frac{1}{N-1}\overset{N}{\underset{i=1}{\sum}}(\bar{x}_{i}-\bar{x})^{2}, ~v=\frac{1}{N}\overset{N}{\underset{i=1}{\sum}}\frac{1}{n_{i}}.
\end{align}

Observe that
\begin{align*}
E[\bar{x}_{i}] & =E[\varphi_{i}]=E[\bar{x}],~(\textrm{where $\varphi_{i}$ is the population mean})\\
E[s_{i}^{2}] & =E[\sigma_{i}^{2}]=E[s^{2}],~(\textrm{where $\sigma_{i}$ is the population variance})\\
Var(\bar{x}_{i}) & =E[Var(\bar{x}_{i}|F_{i})]+Var[E(\bar{x}_{i}|F_{i})]=\frac{E\sigma_{i}^{2}}{n_{i}}+Var(\varphi_{i}),\\
(N-1)u^{2} & =\overset{N}{\underset{i=1}{\sum}}(\bar{x}_{i}-\bar{x})^{2},\\
\textrm{with expectation:}\\
(N-1)Eu^{2}= & {\underset{i}{\sum}}E\bar{x}_{i}^{2}-NE\bar{x}^{2}\\
= & {\underset{i}{\sum}}[Var(\bar{x}_{i})+(E\bar{x}_{i})^{2}]-NE[Var(\frac{{\underset{i}{\sum}}~\bar{x}_{i}}{N})+(E\bar{x})^{2}]\\
= & {\underset{i}{\sum}}(Var(\bar{x}_{i})+(E\bar{x})^{2})-N(\frac{{\underset{i}{\sum}}Var(\bar{x}_{i})}{N^{2}}+(E\bar{x})^{2})),\\
\textrm{so}~~~~~~~~~~Eu^{2}= & \frac{1}{N}\sum~Var(\bar{x}_{i})=\frac{1}{N}\sum(\frac{E\sigma_{i}^{2}}{n_{i}}+Var(\varphi_{i}))\\
= & \frac{1}{N}\sum\frac{1}{n_{i}}(E\sum s_{i}^{2})+Var(\varphi_{i})=vEs^{2}+Var\varphi_{i}.
\end{align*}

We have
\begin{align}
Var~\varphi_{i}= & E(u^{2}-vs^{2}), \nonumber \\
Var~\bar{x}_{i}= & E(u^{2}-vs^{2})+\frac{1}{n_{i}}Es^{2}. \label{eq:Estimation of variances interval}
\end{align}
where the statistics $u^{2},s^{2}$ are single values, we use $(u^{2},vs^{2})^{+}=\textrm{max } \{0,u^{2}-vs^{2}\}$ to estimate $Var~\varphi_{i}$ and $(u^{2}-vs^{2})+\frac{1}{n_{i}}s^{2}$
to estimate $Var~\bar{x}_{i}$.

Combining equation (\ref{eq:Estimation of variances interval}) with equation (\ref{eq:a&B solution interval}), $a$ and $b$ take the form (i=1,...,N)
\begin{align}
b_{i}= & (u^{2}-vs^{2})/(u^{2}-vs^{2}+\frac{s^{2}}{n_{i}}), \nonumber \\
a_{i}= & \bar{x}-b_{i}*\bar{x}.
\end{align}


The linear EB estimator of the symbolic means for the intervals is computed as
\begin{align}
\hat{t_{i}}= a+b*\bar{x}_{i}. \label{eq:Linear EB estimator of symbolic mean}
\end{align}

In the empirical Bayes paradigm, we can think of the data for each individual as coming from different distributions, which means we have $N$ groups with only one observation ($n_{i}=1$) for each distribution.
This can not be done if we are using other type of data.
Interval data contain not only the value of each individual but also have a measure of variability of the value.

To illustrate the methodology, we will use the horse data exhibited in Table \ref{tb:Horse data} which is obtained from the SODAS website.\footnote{\emph{https://www.ceremade.dauphine.fr/SODAS/exemples.htm} which has been removed}
This data set is in interval form, gives the size of each horse.
(We only show the first 10 horses and the intervals to which they belong.)
\begin{table}[H]
\centering %
\begin{tabular}{cc}
\hline
Size(low)  & Size(high)\\
 135 & 147 \\
 130 & 150 \\
 135 & 148 \\
 135 & 147 \\
 145 & 155 \\
 145 & 160 \\
 140 & 157 \\
 150 & 167 \\
 150 & 172 \\
 150 & 170 \\
\hline
\end{tabular}\caption{Horse data}
\label{tb:Horse data}
\end{table}

After computing the a and b, we then apply the linear EB formula (\ref{eq:Linear EB estimator of symbolic mean})
to the symbolic means to get estimates:
\begin{table}[H]
\centering %
\begin{tabular}{cc}
\hline
Symbolic Mean (center)  & EB estimates \\
\hline
141 &142.5985\\
140 &141.7302\\
141.5 &143.0326\\
141 &142.5985\\
150 &150.4128\\
152.5 & 152.5834\\
148.5 & 149.1104\\
158.5 & 157.793\\
161 & 159.9636\\
160 & 159.0954\\
\hline
\end{tabular}\caption{Symbolic means and EB Estimated means of horse data}
\label{tb:Symbolic means and EB Estimated means of horse data}
\end{table}

The results obtained are shown in Table \ref{tb:Symbolic means and EB Estimated means of horse data}. As the overall symbolic mean is 153.1333 while the mean of our estimates is also 153.1333. We can observe that our estimation will make the data move closer to the "mean".\\
%

Next we want to generalize the scalar form method to vector form.
Each vector X is now several interval-valued variables given form $[\boldsymbol{l},\boldsymbol{u}]$, where $\boldsymbol{l}$ and $\boldsymbol{u}$ are vectors which represent the lower and upper bounds of these interval-valued variables.

Similarly, we wish to minimize
\begin{align}
E[((a+BX)-\varphi)((a+BX)-\varphi)^{T}],
\end{align}
where $\varphi$ is the "mean" vector, a is a vector and B is a matrix.

Combining what we have done for the vector form (\ref{eq:Linear EB estimates vector form}) and interval-valued variables (\ref{eq:Linear EB estimator of symbolic mean}), we end up with the estimator as follows:
\begin{align}
\hat{t_{i}}= &~ a+B\bar{X_{i}}, \nonumber \\
a= & ~\bar{X}-B\bar{X}, \nonumber \\
B= & ~(U^{2}-vS^{2})^{+}[(U^{2}-vS^{2})^{+}+\frac{1}{n_{i}}S^{2}]^{-1}.
\end{align}

where the statistics are defined as
\begin{align*}
 & \bar{X_{i}}=\frac{1}{n_{i}}\overset{n_{i}}{{\underset{j=1}{\sum}}}\frac{\boldsymbol{l_{ij}}+\boldsymbol{u_{ij}}}{2}, ~~~~~~~~~~~~~~~~~~~~~~~~~~~~~~~~~~~~~~~~~~~\bar{X}=\frac{1}{N}\overset{N}{\underset{i}{\sum}}\bar{X_{i}},\\
 & S_{i}^{2}=\frac{1}{3n_{i}}\overset{n_{i}}{{\underset{j=1}{\sum}}}(\boldsymbol{u_{ij}}^{2}+\boldsymbol{u_{ij}}\boldsymbol{l_{ij}}+
 \boldsymbol{l_{ij}}^{2})-\frac{1}{4m^{2}}\overset{n_{i}}{{\underset{j=1}{\sum}}}(\boldsymbol{l_{ij}}+\boldsymbol{u_{ij}})^{2},
 ~S^{2}=\frac{1}{N}\overset{N}{\underset{i}{\sum}}S_{i}^{2},\\
 & U^{2}=\frac{1}{N-1}\overset{N}{\underset{i}{\sum}}(\bar{X_{i}}-\bar{X})(\bar{X_{i}}-\bar{X})^{T},
 ~~~~~~~~~~~~~~~~~~~~~~~ v=\frac{1}{N}\overset{N}{\underset{i}{\sum}}\frac{1}{n_{i}}.
\end{align*}

\subsection{Further exploration}

Cluster analysis is a statistical technique that aims at grouping together objects in a number of clusters, based on the observed values for a set of variables.
The constructed clusters are organized according to their similarities or differences.
In this subsection we will discuss the dynamic clustering method based on several distance measures and its application with the empirical
Bayes estimator.

\subsubsection{Algorithm}

Here we use a dynamic clustering algorithm (DCA) first proposed by Diday in 1971 and reorganized by Diday \& Simon in 1976\cite{20}.
We give a brief description of the algorithm.
The dynamic clustering algorithm represents an unsupervised non-hierarchical clustering method which can be proven
to generalize several clustering partition methods such as k-means and k-median algorithm.\cite{33}

Let $E=\{1,...,n\}$ be a set of n symbolic objects with p-dimensions $(X_{1},...,X_{p})$, where
$X_{j}=[l_{j},u_{j}]\in \{[l,u]:~l,u\in \mathbb{R},~l\leqslant u  \}$ for j=1,...,p.
This algorithm searches a partition P = $\{C_{1},C_{2},...,C_{K}\}$ of E in a fixed number K clusters by minimizing a criterion W which evaluates the dissimilarity between the clusters and their representatives.

If each cluster is represented by a prototype, this algorithm also determines a set of p-dimensional
class representatives L = $\{l_{1},l_{2},...,l_{K}\}$.
A good partition places similar observation in a cluster which is close to
the prototype, while dissimilar to the prototype of any other clusters.
Consequently, the clustering criterion minimizes the following function
\begin{align}
W(P,L)=\sum_{h=1}^{K}\delta(C_{h},l_{h}), \label{eq:Dissimilarity criterion}
\end{align}
where $\delta(\cdot,\cdot)$ is a dissimilarity measure between a class $C_{h}$ and a class prototype $l_{h}$.

The dissimilarity measure $\delta(\cdot,\cdot)$ is usually expressed as
\begin{align}
\delta(C_{h},l_{h})=\sum_{x_{i}\in C_{h}}\sum_{j=1}^{p}d(x_{ij},l_{hj}),~~C_{h}\in E,~l_{h}\in L. \label{eq:Dissimilarity measure}
\end{align}
where $d(\cdot,\cdot)$ represents a distance function between two interval-valued variables.
We consider several distances measures as follows:

L2 distance\cite{17}:
\begin{align}
d_{L_{2}}(x_{i},x_{j}) & =({|l_{i}-l_{j}|}^{2}+{|u_{i}-u_{j}|}^{2})^{1/2} \nonumber \\
 & =(2(c_{i}-c_{j})^{2}+2(r_{i}-r_{j})^{2})^{1/2}.
\end{align}

Hausdorff distance\cite{18}:
\begin{align}
d_{Hau}(x_{i},x_{j}) & =max\{|l_{i}-l_{j}|,|u_{i}-u_{j}|\} \nonumber \\
 & =|c_{i}-c_{j}|+|r_{i}-r_{j}|.
\end{align}

Wasserstein distance\cite{28}:
\begin{align}
d_{Wass}(x_{i},x_{j}) & =(\int_{0}^{1} |F^{-1}(t)-G^{-1}(t)|^{2}dt)^{1/2} \nonumber \\
& =((c_{i}-c_{j})^{2}+\frac{1}{3}(r_{i}-r_{j})^{2})^{1/2}.
\end{align}
where $c_{i}=\frac{l_{i}+u_{i}}{2},r_{i}=\frac{u_{i}-l_{i}}{2}$, $F^{-1}$ and $G^{-1}$ are the inverse distribution functions of two random variables which are uniformly distributed on the interval.

The transformation of the distances from bounds to center and half-width included in Appendix A.1.1. The Waseerstein distance is derived from Irpino and Verde (2008)\cite{33}.

Then the algorithm \ref{Algorithm: DCA} iteratively performs a two-stage procedure as follows\cite{32}:

 (a) Given L fixed, finding P that minimizes W(P,L) is obtained by finding the class $C_{h}$,
$ C_{h}=\{i\in E|\overset{p}{\underset{j=1}{\sum}}d(x_{ij},l_{hj})\leqslant \overset{p}{\underset{j=1}{\sum}}d (x_{ij},l_{kj}),\forall k=1,2,...,K\}$ for h = 1,..., K;

(b) Given P fixed, finding L that minimizes W(P,L) is obtained by finding the prototypes $l_{h}$,
$l_{h}=\underset{l\in C_{h}}{argmin}~\delta(C_{h},l)$ for h = 1,..., K. \\
But this procedure often fall to the local optimum, so I used $l_{h}=\underset{l\in C_{h}}{Ave}\{l\}$.

\begin{table}[H]
\centering %
\begin{tabular}{|l|}
\hline
1. Initialization \tabularnewline
~~~~Randomly select a partition $\{C_{1},C_{2},\cdots,C_{K}\}$ of E and K prototypes. \tabularnewline
2. Allocation stage \tabularnewline
~~~~test $\leftarrow$ 0 \tabularnewline
~~~~\textbf{for} i=1 to n do \tabularnewline
~~~~~~~~~~~~Find cluster $C_{h}$ such that \tabularnewline
~~~~~~~~~~~~$h=\underset{k=1,...,K}{argmin}~\overset{p}{\underset{j=1}{\sum}}d(x_{ij},l_{kj})$\\
~~~~~~~~~~~~\textbf{if} $x_{i}\in C_{k}$ and $k\neq h$ then \tabularnewline
~~~~~~~~~~~~~~~~test $\leftarrow$ 1 \tabularnewline
~~~~~~~~~~~~~~~~$C_{h}\leftarrow C_{h}\cup\{i\}$ \tabularnewline
~~~~~~~~~~~~~~~~$C_{k}\leftarrow C_{k}\setminus\{i\}$ \tabularnewline
~~~~~~~~~~~~\textbf{end if} \tabularnewline
~~~~\textbf{end for} \tabularnewline
3. Representative stage \tabularnewline
~~~~\textbf{for} h=1 to K do\tabularnewline
~~~~~~~~~~~~compute the prototype $l_{h}$ \tabularnewline
~~~~\textbf{end for} \tabularnewline
4. Stopping criterion \tabularnewline
~~~~\textbf{if} test=0 then STOP, otherwise return to Step 2.\tabularnewline
\hline
\end{tabular}\caption{Algorithm: DCA}
\label{Algorithm: DCA}
\end{table}

Before applying it to a real-life data set, we may preproccess the data by standardizing in the following ways of paper\cite{17}, to guarantee the variables are on the same scale.

1. Standardization using the dispersion of the interval centers
\begin{align*}
l'_{ij}=\frac{l_{ij}-m_{j}}{s_{ij}}~~~~~~\textrm{and}~~~~~~u'_{ij}=\frac{u_{ij}-m_{j}}{s_{ij}}
\end{align*}
where $m_{j}=\frac{1}{n}\overset{n}{\underset{i=1}{\sum}}\frac{l_{ij}+u_{ij}}{2}$ represents mean of centers
and dispersion $s_{j}^{2}=\frac{1}{n}\overset{n}{\underset{i=1}{\sum}}(\frac{l_{ij}+u_{ij}}{2}-m_{j})^{2}$.

This method standardizes such that the transformed intervals have centers with mean 0 and dispersion 1 in each dimension.

2. Standardization using the dispersion of the interval boundaries.
\begin{align*}
l'_{ij}=\frac{l_{ij}-m_{j}}{\tilde{s_{j}}}~~~~~~\textrm{and}~~~~~~u'_{ij}=\frac{u_{ij}-m_{j}}{\tilde{s_{j}}}
\end{align*}
where $\tilde{s_{j}}^{2}=\frac{1}{n}\overset{n}{\underset{i=1}{\sum}}\frac{(l_{ij}-m_{j})^{2}+(u_{ij}-m_{j})^{2}}{2}$.

This method standardizes such that the mean and the joint dispersion of the transformed interval
boundaries are 0 and 1, respectively.

3. Standardization using the global range.
\begin{align*}
l'_{ij}=\frac{l_{ij}-Min_{j}}{Max_{j}-Min_{j}}~~~~~~\textrm{and}~~~~~~u'_{ij}=\frac{u_{ij}-Min_{j}}{Max_{j}-Min_{j}}
\end{align*}
where $Min_{j}=min\{l_{1j},...,l_{nj}\}$ and $Max_{j}=max\{u_{1j},...,u_{nj}\}$.

The last method transforms the range of the re-scaled intervals to be [0, 1].

Having seen the algorithm and data-preparation for the clustering of interval data, we continue our application.

\begin{description}
\item [{Application}] Blood Pressure data
\end{description}
\addcontentsline{toc}{section}{Application: Blood Pressure data}

The data comes from a project with The Children's Hospital of Eastern Ontario. This dataset recorded three measurements of systolic blood pressure of each patient, so we are able to summarize the data as interval data. We first apply the dynamic clustering algorithm to the original data, to see which points are grouped together. Then we will be able to use our estimator to compute the posterior mean of each patients and use ordinary K-means approach to cluster them.

We will visualize the dynamic clustering results in two dimensions (axis x is the lower bound and axis y is the upper bound).

\begin{figure}[H]
\centering
\includegraphics[width=9.5cm]{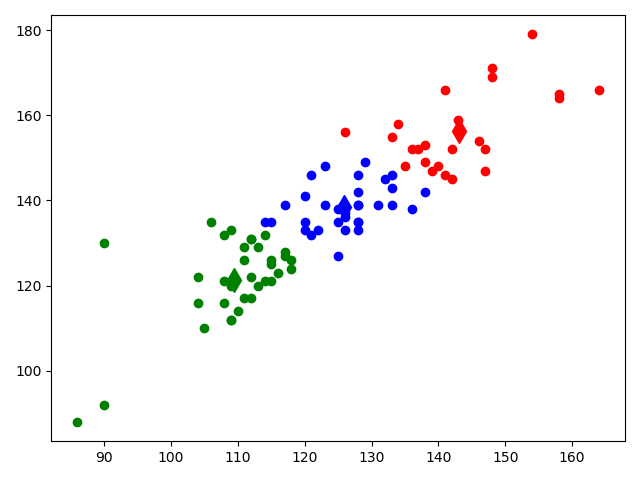}
\caption{Scatter plots of interval clustering}
\label{fg:Scatter plots of interval clustering}
\end{figure}

Figures \ref{fg:Scatter plots of interval clustering} shows that three clusters distinguished. Three different color denote three clusters, green points represent the people whose blood pressure is normal, the blue point represent the people who has predisposition to high blood pressure, red points represent people with hypertension.

\begin{figure}[H]
\centering \includegraphics[width=9.5cm]{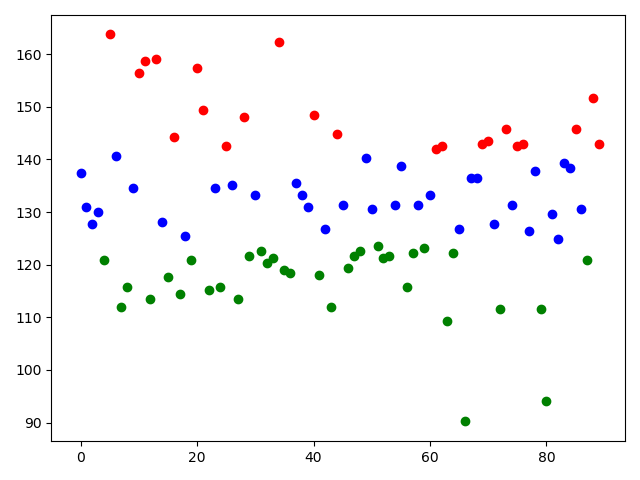}
\caption{Scatter plots of estimated symbolic means}
\label{fg:Scatter plots of estimated symbolic means}
\end{figure}

Then we plot the clustering result of our estimated symbolic means in Figure \ref{fg:Scatter plots of estimated symbolic means}.

The colors represent the points belong to one cluster. The clustering result is consistent with Figure \ref{fg:Scatter plots of interval clustering}, which proves that the EB estimation has captured the characteristics of individuals. By transforming the data into one dimension, it can be more easily clustered and analysed.

\chapter{Tweedie's formula}
\section{Introduction}

In 1956, H. Robbins points out an extraordinary Bayesian estimation formula\cite{43}.
Suppose that x given parameter $\theta$ has density $g_{\theta}(x)$ in what follows and $\theta$ has been
sampled from a prior density $\pi(\theta)$. Tweedie's formula calculates the posterior expectation of $\theta$ given x.
This formula has wide application for exponential families.
The work of Efron (2011)\cite{25} further illustrates the use of Tweedie's formula, which provides
an effective bias correction tool when estimating a great number of parameters within the context of micro-arrays.
In this study, we generalize it to multivariate exponential families and combine it with some ancient theories.

\section{Univariate Version}

We begin with a recap of the univariate version of Tweedie's formula. Suppose we have a random variable x, and its density given
$\theta$ is $x|\theta\sim g_{\theta}(x)$ belonging to one parameter exponential family where:
\begin{align}
g_{\theta}(x)=e^{\theta x-K(\theta)}g_{0}(x),~\theta\in\Theta,~x\in\mathcal{X}
\end{align}
with respect to some measure $m(dx)$ such as the uniform on the real line or a discrete measure on the integers.
The sets $\Theta,\mathcal{X}$ are both subsets of the real line, $\theta$ is the natural or canonical parameter of the family.\cite{2}
$K(\theta)$ is the cumulant function or log of the moment generating function of $g_{0}(x)$ (which makes $g_{\theta}(x)$
integrate to 1), and $g_{0}(x)$ labeled the carrying density.

The normalizing constant $K(\theta)$, computed under $g_{0}(x)$
\begin{align}
e^{K(\theta)} =\int_{x\in \mathcal{X}}g_{0}(x)e^{\theta x}m(dx). \nonumber
\end{align}

$K(\theta)$ is a strictly convex functions such that
\begin{align}
E_{\theta}X=K'(\theta),~Var_{\theta}X=K''(\theta).
\end{align}
where $'$ indicates derivative. $K'(\theta)$ is an increasing function of $\theta$ as $K''(\theta)$ is non-negative.

The one parameter exponential family includes the Normal, Exponential, Gamma, Binomial, Negative Binomial, Poisson distribution.

Suppose we take a Bayesian approach and assume prior $\theta\sim\pi(\theta)$ with respect to Lebesgue measure on $\Theta$. Bayes theorem provides the posterior density of $\theta$ given x:
\begin{align}
g(\theta|x)=\frac{g_{\theta}(x)\pi(\theta)}{g_{X}(x)},
\end{align}
where $g_{X}(x)$ is the marginal distribution of $x$ given by
\begin{align}
g_{X}(x) & =\int{g_{\theta}(x)\pi(\theta)}d\theta
\end{align}

Set
\begin{align}
\lambda(x)&=log(\frac{g_{X}(x)}{g_{0}(x)}), \nonumber \\
\pi_{0}(\theta)&=\pi(\theta)e^{-K(\theta)}.  \nonumber
\end{align}

The posterior density of $\theta$ given $x$ is:
\begin{align}
g(\theta|x)=e^{x\theta-\lambda(x)}\pi_{0}(\theta)
\end{align}
represents once again a member of the exponential family with x as natural (canonical) parameter and $\lambda(x)$ as the new cumulant function.

Due to the property of exponential families, differentiating $\lambda(x)$ yields the Tweedie's formula which computes posterior
mean and variance of $\theta$ given x as
\begin{align}
E(\theta|x) & =\lambda'(x)=\frac{g_{X}'(x)}{g_{X}(x)}-\frac{g_{0}'(x)}{g_{0}(x)},   \\
Var(\theta|x) & =\lambda''(x)=[\frac{g_{X}'(x)}{g_{X}(x)}]'-[\frac{g_{0}'(x)}{g_{0}(x)}]'.
\end{align}
and similarly $Skewness[\theta|x] = \lambda'''(x)/\lambda''(x)^{\frac{3}{2}}$.
The literature has not shown much interest in the higher moments of $\theta$ given x, but they can be derived naturally from exponential family.\cite{21}

Generally, the quantity $g_{X}(x)$ which depends on the marginal distribution, is unknown if the prior $\pi(\theta)$ on $\theta$ is not specified.
However, in the empirical Bayes paradigm, the density $g_{X}(x)$ and its derivative $g_{X}'(x)$ can be estimated from the data. In this study, we propose the use of the Pearson system of distributions to approximate the marginal distribution which we will introduce in a later section.
The quantity $g_{0}(x)$ and its derivative are independent of the prior and can be calculated directly from the distribution.

\subsection{Normal distribution}
We demonstrate how to use the Tweedie's formula to compute empirical Bayes estimates in a normal distribution.
\subsubsection{$\sigma^{2}$ known}

Let $x_{1},...,x_{N}$ be a random sample from $N(\mu,\sigma^{2})$.
In this case $\theta=\frac{\mu}{\sigma^{2}}$ and
\begin{align}
g_{0}(x)=\frac{1}{\sqrt{2\pi\sigma^{2}}}e^{-\frac{x^{2}}{2\sigma^{2}}}. \nonumber
\end{align}

When $\sigma^{2}$ is known, we can directly plug in Tweedie's formula,
that the posterior mean of $\mu$ conditional on $x$ is given by
\begin{align}
E\left[\mu|x\right]=x+\sigma^{2}\left(\frac{g_{X}'\left(x\right)}{g_{X}\left(x\right)}\right), \label{eq:Tweedie normal mean}
\end{align}
where the marginal density of $X$ is $g_{X}\left(x\right).$

The first term $x$ comes from the derivative of $g_{0}(x)$. The term $\sigma^{2}\frac{g_{X}'(x)}{g_{X}(x)}$
for the posterior mean is the Bayes bias correction to
the maximum likelihood estimate x. $\sigma^{2}$ is known and $\frac{g_{X}'(x)}{g_{X}(x)}$ are estimated immediately from \{$x_{1},\cdots ,x_{N}$\} using the Pearson system.

\subsubsection{$\sigma^{2}$ unknown}

Now we wish to consider the case where $\sigma^{2}$ is unknown. The discussion below is original.
We know that for the normal distribution, the sample mean is independent of the sample variance. Consequently, we consider the variable
\begin{align}
w=\frac{\nu s^{2}}{\sigma^{2}}\sim\chi_{\nu}^{2},
\end{align}
where $\nu=n-1$ and
\begin{align}
s^{2}=\frac{\sum\left(x_{i}-\bar{x}\right)^{2}}{\nu}. \nonumber
\end{align}

The density of $u\equiv s^{2}$ is therefore
\begin{align}
f\left(u\right)=\left[\varGamma\left(\nu/2\right)\right]^{-1}\left(\frac{\nu}{\sigma^{2}}\right)^{\nu/2}u^{\left(\nu/2\right)-1}e^{-\frac{\nu u}{2\sigma^{2}}},u>0
\end{align}
where we now define
\begin{align}
c=\left[\varGamma\left(\nu/2\right)\right]^{-1}\left(\nu\right)^{\nu/2} \nonumber
\end{align}

Place a prior $\pi\left(\sigma^{2}\right)$ and then the marginal density of $u$ is
\begin{align}
h(u)=c\int u^{(\nu/2)-1}e^{-\frac{\nu u}{2\sigma^{2}}}(\frac{1}{\sigma^{2}})^{\nu/2}\pi(\sigma^{2})d\sigma^{2}
\end{align}

It follows
\begin{align}
h'\left(u\right) & =  \frac{d}{du}c\int u^{\left(\nu/2\right)-1}\left(\sigma^{2}\right)^{-\nu/2}e^{-\frac{\nu u}{2\sigma^{2}}}\pi\left(\sigma^{2}\right)d\sigma^{2} \nonumber \\
 & =  \left(\frac{\nu}{2}-1\right)\frac{1}{u}h\left(u\right)-\frac{\nu}{2}h\left(u\right)E\left[\frac{1}{\sigma^{2}}|u\right] \nonumber \\
\frac{h'\left(u\right)}{h\left(u\right)} & =\left(\frac{\nu}{2}-1\right)\frac{1}{u}  -\frac{\nu}{2}E\left[\sigma^{-2}|u\right].
\end{align}

We may now obtain the posterior mean of $\sigma^{2}$ by means of
a Taylor series around $u$, namely
\begin{align}
\frac{1}{\sigma^{2}}=\frac{1}{u}-\frac{1}{u^{2}}\left(\sigma^{2}-u\right)
\end{align}

Taking conditional expectation we get
\begin{align}
E\left[\sigma^{-2}|u\right] & =  \frac{1}{u}-\frac{1}{u^{2}}\left(E\left[\sigma^{2}|u\right]-u\right) \nonumber \\
 & =  \frac{2}{u}-\frac{1}{u^{2}}E\left[\sigma^{2}|u\right], \nonumber
\end{align}
which leads to
\begin{align}
E\left[\sigma^{2}|u\right] & =  2u-u^{2}E\left[\sigma^{-2}|u\right] \nonumber \\
 & =  2u-u^{2}\left\{ \frac{2}{\nu}\left(\frac{\nu}{2}-1\right)\frac{1}{u}-\frac{2}{\nu}\frac{h'\left(u\right)}{h\left(u\right)}\right\} \nonumber \\
 & =  2u-u\left(1-\frac{2}{\nu}\right)+\frac{2}{\nu}u^{2}\frac{h'\left(u\right)}{h\left(u\right)} \nonumber \\
 & =  u\left(1+\frac{2}{\nu}\right)+\frac{2}{\nu}u^{2}\frac{h'\left(u\right)}{h\left(u\right)}. \label{eq:Normal variance estimator}
\end{align}

If $\nu$ is large, the middle term $\frac{2u}{\nu}$ is negligible and hence the quantity
$\frac{2}{\nu}u^{2}\frac{h'(u)}{h(u)}$ is the
bias correction. This term can be approximated by the Pearson system. We do not pursue that in this thesis.


\section{Multivariate Version}

Our contribution in this section is to generalize Tweedie's formula into multivariate version and derived the formula for two specific multivariate distributions.

Suppose we have a random vector $\mathbf{x}=(x_{1},...,x_{k})^{T}$,
and its distribution given $\boldsymbol{\theta}=(\theta_{1},...,\theta_{k})^{T}$ belongs to multi parameter exponential family having density:
\begin{align}
g_{\boldsymbol{\theta}}(\mathbf{x})=exp\{\boldsymbol{\theta}^{T}\boldsymbol{x}-K(\boldsymbol{\theta})\}g_{0}(\boldsymbol{x}),
~\boldsymbol{\theta}\in\boldsymbol{\Theta},~\boldsymbol{x}\in\boldsymbol{\mathcal{X}}
\end{align}
with $\boldsymbol{\theta}, \mathbf{x}$ are now vectors and $\boldsymbol{\theta}\sim\pi(\boldsymbol{\theta})$ has an unknown prior density.

$g_{0}(\boldsymbol{x})$ is the carrying density, the function $K(\theta)$ is the cumulant function computed under $g_{0}(\boldsymbol{x})$
\begin{align*}
K(\boldsymbol{\theta})=logE(exp(\boldsymbol{\theta}^{T}\boldsymbol{x})).
\end{align*}

According to Bayes theorem, the posterior density of $\boldsymbol{\theta}$ is
given by:
\begin{align}
g(\boldsymbol{\theta}|\boldsymbol{x})&=\frac{g_{\boldsymbol{\theta}}(\boldsymbol{x})\pi(\boldsymbol{\theta})}{g_{X}{(\boldsymbol{x})}}  \nonumber \\
&=e^{x\boldsymbol{\theta}-\lambda(\boldsymbol{x})}\pi_{0}(\boldsymbol{\theta}).
\end{align}
where $g_{X}(\boldsymbol{x})$ is the marginal distribution of $\mathbf{x}$
\begin{align*}
g_{X}(\boldsymbol{x}) & =\int{g_{\boldsymbol{\theta}}(\boldsymbol{x})\pi(\boldsymbol{\theta})}d\boldsymbol{\theta}
\end{align*}

This represents an exponential family with $\mathbf{x}$ is natural (canonical) parameter,
$\lambda(\boldsymbol{x})=log(\frac{g_{X}(\boldsymbol{x})}{g_{0}(\boldsymbol{x})})$ is the cumulant function
(normalizing constant) and $\pi_{0}(\boldsymbol{\theta})=\pi(\boldsymbol{\theta})e^{-K(\boldsymbol{\theta})}$ is
the carrying density.

Differentiating $\lambda(\boldsymbol{x})$ yields the posterior cumulants of $\boldsymbol{\theta}$ given $\mathbf{x}$,
\begin{align}
E(\boldsymbol{\theta}|\boldsymbol{x})=\lambda'(\boldsymbol{x})~~,~~~~Var(\boldsymbol{\theta}|\boldsymbol{x})=\lambda''(\boldsymbol{x}).
\end{align}

Letting
\begin{align}
l(\boldsymbol{x})&=log(g_{X}(\boldsymbol{x})), \nonumber \\
l_{0}(\boldsymbol{x})&=log(g_{0}(\boldsymbol{x})).
\end{align}

We can express the posterior mean and variance of $\theta|x$ as
\begin{align}
E(\boldsymbol{\theta}|\boldsymbol{x}) & =l'(\boldsymbol{x})-l_{0}'(\boldsymbol{x}), \\
Var(\boldsymbol{\theta}|\boldsymbol{x}) & =l''(\boldsymbol{x})-l_{0}''(\boldsymbol{x}).
\end{align}

Or we can write as
\begin{align}
E(\theta_{j}|\boldsymbol{x}) & =\frac{1}{g_{X}(\boldsymbol{x})}\frac{\partial g_{X}(\boldsymbol{x})}{\partial x_{j}}-\frac{1}{g_{0}(\boldsymbol{x})}\frac{\partial g_{0}(\boldsymbol{x})}{\partial x_{j}},\\
Var(\theta_{j}|\boldsymbol{x}) & =\frac{\partial}{\partial x_{j}}[\frac{1}{g_{X}(\boldsymbol{x})}\frac{\partial g_{X}(\boldsymbol{x})}{\partial x_{j}}]-\frac{\partial}{\partial x_{j}}[\frac{1}{g_{0}(\boldsymbol{x})}\frac{\partial g_{0}(\boldsymbol{x})}{\partial x_{j}}],
\end{align}
where $\theta_{j}$ is a component of the vector.

\subsection{Multivariate Normal Distribution}

Consider Gaussian distribution, $\mathbf{x}|\boldsymbol{\mu}\sim N_{p}(\boldsymbol{\mu},\Sigma)$
\begin{align*}
g_{\boldsymbol{\mu}}(\boldsymbol{x}) & =\frac{1}{\sqrt{(2\pi)^{p}|\Sigma|}}exp\{-\frac{1}{2}(\boldsymbol{x}-\boldsymbol{\mu})^{T}\Sigma^{-1}(\boldsymbol{x}-\boldsymbol{\mu})\},\\
g_{0}(\boldsymbol{x}) & =\frac{1}{\sqrt{(2\pi)^{p}|\Sigma|}}exp\{-\frac{1}{2}\boldsymbol{x}^{T}\Sigma^{-1}\boldsymbol{x}\}.
\end{align*}
We have $K(\boldsymbol{\mu})=\frac{1}{2}\boldsymbol{\mu}^{T}\Sigma^{-1}\boldsymbol{\mu}$.

\subsubsection{Posterior mean}

We consider taking derivative with the integral, notice that
\begin{align*}
g_{X}(\boldsymbol{x}) & =\int{g_{\boldsymbol{\mu}}(\boldsymbol{x})\pi(\boldsymbol{\mu)}}d\boldsymbol{\mu}=\int{e^{\boldsymbol{\mu}^{T}\Sigma^{-1}\boldsymbol{x}-K(\boldsymbol{\mu})}g_{0}(\boldsymbol{x})\pi(\boldsymbol{\mu)}}d\boldsymbol{\mu}\\
 & =\int{e^{\boldsymbol{\mu}^{T}\Sigma^{-1}\boldsymbol{x}-\frac{1}{2}\boldsymbol{\mu}^{T}\Sigma^{-1}\boldsymbol{\mu}}\frac{1}{\sqrt{(2\pi)^{p}|\Sigma|}}e^{-\frac{1}{2}\boldsymbol{x}^{T}\Sigma^{-1}\boldsymbol{x}}\pi(\boldsymbol{\mu})}d\boldsymbol{\mu}\\
 & =\int{e^{\boldsymbol{\mu}^{T}\Sigma^{-1}\boldsymbol{x}-\frac{1}{2}\boldsymbol{x}^{T}\Sigma^{-1}\boldsymbol{x}}\frac{1}{\sqrt{(2\pi)^{p}|\Sigma|}}e^{-\frac{1}{2}\boldsymbol{\mu}^{T}\Sigma^{-1}\boldsymbol{\mu}}\pi(\boldsymbol{\mu})}d\boldsymbol{\mu}\\
g_{X}'(\boldsymbol{x}) & =\Sigma^{-1}\int{(\boldsymbol{\mu}-\boldsymbol{x})e^{\boldsymbol{\mu}^{T}\Sigma^{-1}\boldsymbol{x}-\frac{1}{2}\boldsymbol{x}^{T}\Sigma^{-1}\boldsymbol{x}}\frac{1}{\sqrt{(2\pi)^{p}|\Sigma|}}e^{-\frac{1}{2}\boldsymbol{\mu}^{T}\Sigma^{-1}\boldsymbol{\mu}}\pi(\boldsymbol{\mu})}d\boldsymbol{\mu}\\
 & =\Sigma^{-1}\int{\boldsymbol{\mu}e^{\boldsymbol{\mu}^{T}\Sigma^{-1}\boldsymbol{x}-\frac{1}{2}\boldsymbol{\mu}^{T}\Sigma^{-1}\boldsymbol{\mu}}\frac{1}{\sqrt{(2\pi)^{p}|\Sigma|}}e^{-\frac{1}{2}\boldsymbol{x}^{T}\Sigma^{-1}\boldsymbol{x}}\pi(\boldsymbol{\mu})}d\boldsymbol{\mu}\\
 & ~~~-\Sigma^{-1}\int{\boldsymbol{x}e^{\boldsymbol{\mu}^{T}\Sigma^{-1}\boldsymbol{x}-\frac{1}{2}\boldsymbol{\mu}^{T}\Sigma^{-1}\boldsymbol{\mu}}\frac{1}{\sqrt{(2\pi)^{p}|\Sigma|}}e^{-\frac{1}{2}\boldsymbol{x}^{T}\Sigma^{-1}\boldsymbol{x}}\pi(\boldsymbol{\mu})}d\boldsymbol{\mu}\\
 & =\Sigma^{-1}\int{\boldsymbol{\mu}\frac{g_{\boldsymbol{\mu}}(\boldsymbol{x})\pi(\boldsymbol{\mu})}{g_{X}(\boldsymbol{x})}}d\boldsymbol{\mu}\cdot g_{X}(\boldsymbol{x})-\Sigma^{-1}\boldsymbol{x}\int{g_{\boldsymbol{\mu}}(\boldsymbol{x})\pi(\boldsymbol{\mu)}}d\boldsymbol{\mu}\\
\frac{g_{X}'(\boldsymbol{x})}{g_{X}(\boldsymbol{x})} & =\Sigma^{-1}E(\boldsymbol{\mu}|\boldsymbol{x})-\Sigma^{-1}\boldsymbol{x}
\end{align*}

Hence,
\begin{align}
E(\boldsymbol{\mu}|\boldsymbol{x})=\boldsymbol{x}+\Sigma l'(\boldsymbol{x}). \label{eq:Posterior mean of multivariate normal}
\end{align}

\subsubsection{Posterior variance}

Continue to derivation leads to
\begin{align*}
g_{X}''(\boldsymbol{x}) & =[g_{X}'(\boldsymbol{x})]'=\Sigma^{-1}\int{[(\boldsymbol{\mu}-\boldsymbol{x})e^{\boldsymbol{\mu}^{T}\Sigma^{-1}\boldsymbol{x}-\frac{1}{2}\boldsymbol{x}^{T}\Sigma^{-1}\boldsymbol{x}}]'\frac{1}{\sqrt{(2\pi)^{p}|\Sigma|}}e^{-\frac{1}{2}\boldsymbol{\mu}^{T}\Sigma^{-1}\boldsymbol{\mu}}\pi(\boldsymbol{\mu})}d\boldsymbol{\mu}\\
 & =\Sigma^{-1}\int{[(\boldsymbol{\mu}^{T}\Sigma^{-1}-\Sigma^{-1}\boldsymbol{x})\boldsymbol{\mu}]e^{\boldsymbol{\mu}^{T}\Sigma^{-1}\boldsymbol{x}-\frac{1}{2}\boldsymbol{x}^{T}\Sigma^{-1}\boldsymbol{x}}\frac{1}{\sqrt{(2\pi)^{p}|\Sigma|}}e^{-\frac{1}{2}\boldsymbol{\mu}^{T}\Sigma^{-1}\boldsymbol{\mu}}\pi(\boldsymbol{\mu})}d\boldsymbol{\mu}\\
 & ~~~~-\Sigma^{-1}\int{[(\boldsymbol{\mu}^{T}\Sigma^{-1}-\Sigma^{-1}\boldsymbol{x})\boldsymbol{x}+1]e^{\boldsymbol{\mu}^{T}\Sigma^{-1}\boldsymbol{x}-\frac{1}{2}\boldsymbol{x}^{T}\Sigma^{-1}\boldsymbol{x}}\frac{1}{\sqrt{(2\pi)^{p}|\Sigma|}}e^{-\frac{1}{2}\boldsymbol{\mu}^{T}\Sigma^{-1}\boldsymbol{\mu}}\pi(\boldsymbol{\mu})}d\boldsymbol{\mu}\\
 & =\Sigma^{-1}\big[\Sigma^{-1}\int{\boldsymbol{\mu}^{T}\boldsymbol{\mu}e^{\boldsymbol{\mu}^{T}\Sigma^{-1}\boldsymbol{x}-\frac{1}{2}\boldsymbol{x}^{T}\Sigma^{-1}\boldsymbol{x}}\frac{1}{\sqrt{(2\pi)^{p}|\Sigma|}}e^{-\frac{1}{2}\boldsymbol{\mu}^{T}\Sigma^{-1}\boldsymbol{\mu}}\pi(\boldsymbol{\mu})}d\boldsymbol{\mu}\\
 & ~~~~-2\Sigma^{-1}\boldsymbol{x}\int{\boldsymbol{\mu}e^{\boldsymbol{\mu}^{T}\Sigma^{-1}\boldsymbol{x}-\frac{1}{2}\boldsymbol{x}^{T}\Sigma^{-1}\boldsymbol{x}}\frac{1}{\sqrt{(2\pi)^{p}|\Sigma|}}e^{-\frac{1}{2}\boldsymbol{\mu}^{T}\Sigma^{-1}\boldsymbol{\mu}}\pi(\boldsymbol{\mu})}d\boldsymbol{\mu}\big]\\
 & ~~~~+\Sigma^{-1}\int{\boldsymbol{x}^{T}\boldsymbol{x}e^{\boldsymbol{\mu}^{T}\Sigma^{-1}\boldsymbol{x}-\frac{1}{2}\boldsymbol{x}^{T}\Sigma^{-1}\boldsymbol{x}}\frac{1}{\sqrt{(2\pi)^{p}|\Sigma|}}e^{-\frac{1}{2}\boldsymbol{\mu}^{T}\Sigma^{-1}\boldsymbol{\mu}}\pi(\boldsymbol{\mu})}d\boldsymbol{\mu}\\
 & ~~~~-\int{e^{\boldsymbol{\mu}^{T}\Sigma^{-1}\boldsymbol{x}-\frac{1}{2}\boldsymbol{x}^{T}\Sigma^{-1}\boldsymbol{x}}\frac{1}{\sqrt{(2\pi)^{p}|\Sigma|}}e^{-\frac{1}{2}\boldsymbol{\mu}^{T}\Sigma^{-1}\boldsymbol{\mu}}\pi(\boldsymbol{\mu})}d\boldsymbol{\mu}\big]\\
 & =\Sigma^{-1}[\Sigma^{-1}E(\boldsymbol{\mu}^{T}\boldsymbol{\mu}|\boldsymbol{x})g_{X}(\boldsymbol{x})-2\Sigma^{-1}\boldsymbol{x}E(\boldsymbol{\mu}|\boldsymbol{x})g_{X}(\boldsymbol{x})+\Sigma^{-1}\boldsymbol{x}^{T}\boldsymbol{x}g_{X}(\boldsymbol{x})-g_{X}(\boldsymbol{x})]\\
\frac{g_{X}''(\boldsymbol{x})}{g_{X}(\boldsymbol{x})}\Sigma^{T} & =\Sigma^{-1}E(\boldsymbol{\mu}^{T}\boldsymbol{\mu}|\boldsymbol{x})-2\Sigma^{-1}\boldsymbol{x}E(\boldsymbol{\mu}|\boldsymbol{x})+\Sigma^{-1}\boldsymbol{x}^{T}\boldsymbol{x}-1\\
 & =\Sigma^{-1}Var[\boldsymbol{\mu}|\boldsymbol{x}]+l'(\boldsymbol{x})(l'(\boldsymbol{x}))^{T}\Sigma^{T}-1
\end{align*}

As
\begin{align*}
l''(\boldsymbol{x}) & =log(g_{X}(x))''=[\frac{g_{X}'(x)}{g_{X}(x)}]'=\frac{g_{X}''(y)g_{X}(x)-g_{X}'(x)(g_{X}'(x))^{T}}{g_{X}(y)g_{X}(x)^{T}}\\
 & =\frac{g_{X}''(\boldsymbol{x})}{g_{X}(\boldsymbol{x})}-l'(\boldsymbol{x})(l'(\boldsymbol{x}))^{T}
\end{align*}

We get
\begin{align}
Var(\boldsymbol{\mu}|\boldsymbol{x})=\Sigma(1+l''(\boldsymbol{x})\Sigma^{T})=\Sigma+\Sigma l''(\boldsymbol{x})\Sigma.
\label{eq:Posterior varianece of multivariate normal}
\end{align}

\subsection{Multinomial Distribution}

\subsubsection{Binomial Case}
We begin with binomial distribution. Suppose that $x\sim Binomial(n,p)$.
Noting that
\begin{align*}
g(x) =\binom{n}{x}p^{x}(1-p)^{n-x}.
\end{align*}

Rewrite in the form of an exponential family
\begin{align*}
g_{\theta}(x) & =2^{n}exp(\theta x-nK(\theta))\big[\frac{\binom{n}{x}}{2^{n}}\big],
\end{align*}

We have the following:
\begin{align*}
g_{0}\left(x\right) & =  \frac{\binom{n}{x}}{2^{n}},\\
\theta & = log\left(\frac{p}{1-p}\right),\\
K\left(\theta\right) & =  log\left(1-p\right) = -log\left(1+exp\theta\right).
\end{align*}

With the notation in Tweedie's formula
\begin{align*}
l_{0}\left(x\right) & = log\,g_{0}\left(x\right), \nonumber \\
l_{0}'\left(x\right) & = \frac{g_{0}'\left(x\right)}{g_{0}\left(x\right)}.
\end{align*}
where
\begin{align}
l_{0}\left(x\right) & =log \frac{n!}{x!(n-x)!}+log\frac{1}{2^{n}} \nonumber \\
 & = -nlog2+logn!-logx!-log\left(n-x\right)!. \label{eq:l0 of binomial}
\end{align}

Compute the marginals
\begin{align*}
g_{X}\left(x\right) & = \int2^{n}exp\left(x\theta-nK\left(\theta\right)\right)g_{0}\left(x\right)\pi\left(\theta\right)d\theta\\
g_{X}'\left(x\right) & = \int x2^{n}exp\left(x\theta-nK\left(\theta\right)\right)g_{0}\left(x\right)\pi\left(\theta\right)d\theta\\
 &~~+ \int2^{n}exp\left(x\theta-nK\left(\theta\right)\right)g_{0}'\left(x\right)\pi\left(\theta\right)d\theta\\
 & = g_{X}\left(x\right)E\left(\theta|x\right)
 +l_{o}'\left(x\right)\int2^{n}exp\left(x\theta-nK\left(\theta\right)\right)g_{0}\left(x\right)\pi\left(\theta\right)d\theta
\end{align*}

Now calculate $l_{o}'(x)$ using the approximation for factorials provided by Stirling's formula
\begin{align}
x!\approx\sqrt{2\pi x}(\frac{x}{e})^{x}=\sqrt{2\pi}x^{x+\frac{1}{2}}e^{-x}. \label{eq:Stirling's formula}
\end{align}

Taking derivative of the logarithm:
\begin{align}
\frac{d}{dx}log\left(x!\right) & \approx log\sqrt{2\pi}+logx+\frac{1}{2x}, \nonumber \\
\frac{d}{dx}log\left(\left(n-x\right)!\right) & \approx log\sqrt{2\pi}-log\left(n-x\right)-\frac{1}{2\left(n-x\right)}.
\label{eq:Approximation of derivative of logs}
\end{align}

Thus plug equation (\ref{eq:Approximation of derivative of logs}) into equation (\ref{eq:l0 of binomial}), we obtain an approximation for the log-marginal distribution
\begin{align}
l_{o}'\left(x\right) & \approx -\left(logx+\frac{1}{2x}\right)+log\left(n-x\right)+\frac{1}{2\left(n-x\right)} \nonumber \\
 & = log\left(\frac{n-x}{x}\right)+\frac{n-2x}{2x\left(n-x\right)}.  \label{eq:Approximation of first derivative of the log-marginal for bonomial}
\end{align}

Substituting equation (\ref{eq:Approximation of first derivative of the log-marginal for bonomial}) into Tweedie's formula and using $(x_{1},...,x_{N})$
to obtain the marginal distribution $g_{X}(x)$ and its derivative,
we are able to compute the posterior mean:
\begin{align}
E\left(\theta|x\right)=\frac{g_{X}'\left(x\right)}{g_{X}\left(x\right)}-l_{o}'\left(x\right). \label{eq:Tweedi's formula for the binomial}
\end{align}

Similarly, we can compute the posterior variance
\begin{align}
Var\left(\theta|x\right)=\frac{g_{X}''(x)}{g_{X}(x)}-[\frac{g_{X}'(x)}{g_{X}(x)}]^{2}-l_{o}''(x).
\end{align}
with the approximated second derivative of log-marginal distribution
\begin{align}
l''_{0}(x)=-\frac{n}{x(n-x)}-\frac{n^{2}-2xn+2x^{2}}{2x^{2}(n-x)^{2}}.
\end{align}

\subsubsection{Multinomial Case}

Suppose that $\mathbf{x} \sim mulnomial(n,p)$.
\begin{align*}
g_{\theta}(\boldsymbol{x}) & =\frac{n!}{x_{1}!\cdots x_{k}!}\prod_{j=1}^{k}p_{j}^{x_{j}}\\
 & =\frac{n!}{x_{1}!...x_{k}!}\prod_{j=1}^{k}exp(x_{j}log(p_{j}))\\
 & =exp(\sum_{j=1}^{k-1}x_{j}\log{(\frac{p_{j}}{1-\sum_{j=1}^{k-1}p_{j}})}+n\log{(1-\sum_{j=1}^{k-1}p_{j})})\frac{n!}{x_{1}!x_{2}!\cdots(n-\sum_{j=1}^{k-1}x_{j})!}.
\end{align*}

Natural parameter is
\begin{align*}
\theta_{j}=\log{(\frac{p_{j}}{1-\sum_{j=1}^{k-1}p_{j}})}.
\end{align*}

The cumulant function is
\begin{align*}
-log(1-\sum_{j=1}^{k-1}p_{j})=log(\sum_{j=1}^{k}e^{\theta_{j}}).
\end{align*}

Under the null hypothesis, $H_{0}$: $\theta_{j}$ = 0, we have that $p_{j}=0$
\begin{align*}
l_{0}(\boldsymbol{x}) & =log(\frac{n!}{x_{1}!x_{2}!\cdots(n-\sum_{j=1}^{k-1}x_{j})!})\\
 & =logn!-log(x_{1}!)-log(x_{2}!)-\cdots-log((n-\sum_{j=1}^{k-1}x_{j})!).
\end{align*}

Again, using Stirling's approximation (\ref{eq:Stirling's formula}),
\begin{align*}
l_{0}(\boldsymbol{x})  & \approx logn!-klog(\sqrt{2\pi})-\sum_{j=1}^{k-1}[(x_{j}+\frac{1}{2})log(x_{j})-x_{j}] \\
 & ~~-[(n-\sum_{j=1}^{k-1}x_{j}+\frac{1}{2})log((n-\sum_{j=1}^{k-1}x_{j}))-(n-\sum_{j=1}^{k-1}x_{j})].
\end{align*}

Hence, its partial derivative will be
\begin{align}
\frac{\partial l_{0}(\boldsymbol{x})}{\partial x_{j}}
& =-(logx_{j}+\frac{1}{2x_{j}})+log(n-\sum_{j=1}^{k-1}x_{j})+\frac{1}{2(n-\sum_{j=1}^{k-1}x_{j})} \nonumber \\
& =log(\frac{n-\sum_{j=1}^{k-1}x_{j}}{x_{j}})+\frac{n-\sum_{j=1}^{k-1}x_{j}-x_{j}}{2x_{j}(n-\sum_{j=1}^{k-1}x_{j})}.
\end{align}

So we obtain the posterior mean for the multinomial distribution for each component
\begin{align}
E(\theta_{j}|x)=\frac{\frac{\partial g_{X}(\boldsymbol{x})}{\partial x_{j}}}{g_{X}(\boldsymbol{x})}-\frac{\partial l_{0}(\boldsymbol{x})}{\partial x_{j}}.
\end{align}

\section{Approximating the marginal distribution}

In this section, we begin by describing the Pearson system of distributions.
Karl Pearson introduced a differential equation to describe a wide variety of distribution functions. This system estimates a distribution from knowledge of the first four moments, easily obtained from the observed data.
We creatively make use of this equation in the context of
Tweedie's formula to approximate the marginal distribution.
On the other hand, Efron (2011)\cite{25} used Tweedie's formula in the context of micro-arrays in order to flag interested genes in prostrate cancer patients. He made use of Lindsey's method which requires a large sample size.
We will show how the use of the Pearson system facilitates the implementation of the empirical Bayes approach.

\subsection{The Pearson system of distributions}

K. Pearson (1895)\cite{41} devised a system to estimate many univariate distributions in terms of four parameters namely
\begin{align}
\frac{1}{f(x)}\frac{df(x)}{dx}=\frac{x-a}{c_{0}+c_{1}x+c_{2}x^{2}}. \label{eq:Pearson system}
\end{align}

The derivation of the Pearson parameters $a,c_{0},c_{1},c_{2}$ comes from Kendall (1948)\cite{37}. Consider certain general results that hold for all members of the family:
\begin{align}
(c_{0}+c_{1}x+c_{2}x^{2})df(x)&=(x-a)f(x)dx \nonumber \\
x^{n}(c_{0}+c_{1}x+c_{2}x^{2})\frac{df}{dx}dx&=x^{n}(x-a)f(x)dx. \nonumber
\end{align}

Integrating both sides over the range of the distribution (assuming that the integrals exist),
\begin{align}
\int_{-\infty}^{\infty}x^{n}(c_{0}+c_{1}x+c_{2}x^{2})f^{\prime}dx=\int_{-\infty}^{\infty}x^{n}(x-a)fdx, \nonumber
\end{align}

Use integration by parts on the left-hand side
\begin{align*}
[x^{n}(c_{0}+c_{1}x+c_{2}x^{2})f]_{-\infty}^{\infty}-\int_{-\infty}^{\infty}[nc_{0}x^{n-1}+(n+1)c_{1}x^{n}+(n+2)c_{2}x^{n+1})]fdx \\
=\int_{-\infty}^{\infty}x^{n+1}fdx-a\int_{-\infty}^{\infty}x^{n}fdx,
\end{align*}

Let us assume that the expression in square brackets vanishes at the
extremities of the distribution,
i.e. $\underset{x\rightarrow\pm\infty}{\lim}x^{n+2}f\rightarrow0$
if the range is infinite.
We then substitute moments for integrals:
\begin{align}
-nc_{0}\mu_{n-1}'-(n+1)c_{1}\mu_{n}'-(n+2)c_{2}\mu_{n+1}'=\mu_{n+1}'-a\mu_{n}'. \label{eq:Recurrence of moments}
\end{align}

Equation (\ref{eq:Recurrence of moments}) gives recurrence relations of origin moments $\mu'_{i+1}$. This permits the determination of any moment from those of lower orders if the constants are known. So conversely all moments can be expressed in terms of $a,c_{0},c_{1},c_{2}$,
$\mu_{0}$(=1) and $\mu_{1}'$. We have four unknown parameters $a,c_{0},c_{1},c_{2}$, put $\mu_{1}'=0$ and n=0, 1, 2, 3 successively in the formula to solve the equations. As central moments are easy to compute in practice and more meaningful in statistics, we express the solutions in terms of the first four central moments:
\begin{align}
c_{0} & =-\frac{\mu_{3}(\mu_{4}+3\mu_{2}^{2})}{A},\nonumber \\
c_{1} & =~a~=-\frac{\mu_{2}(4\mu_{2}\mu_{4}-3\mu_{3}^{2})}{A},\nonumber \\
c_{2} & =-\frac{(2\mu_{2}\mu_{4}-3\mu_{3}^{2}-6\mu_{2}^{2})}{A},\nonumber \\
\textrm{where}~A & =10\mu_{2}\mu_{4}-12\mu_{3}^{2}-18\mu_{2}^{2}.
\end{align}

The four central moments can also be uniquely determined by mean ($\mu_{1}$), variance ($\mu_{2}$), skewness ($\beta_{1}$), and kurtosis ($\beta_{2}$), which are more commonly used parameters for a distribution and easily obtained from statistical tools. They give a good representative of the distribution.

Skewness and Kurtosis are defined as
\begin{align}
\beta_{1}&=E(\frac{X-\mu_{1}}{\sqrt{\mu_{2}}})^{3}=\frac{\mu_{3}}{\mu_{2}^{\frac{3}{2}}} \label{eq:Skewness} \\
\beta_{2}&=E(\frac{X-\mu_{1}}{\sqrt{\mu_{2}}})^{4}=\frac{\mu_{4}}{\mu_{2}^{2}}           \label{eq:Kurtosis}
\end{align}

For a normal distribution $N(\mu_{1},\mu_{2})$, $\beta_{1}$=0 and $\beta_{2}$=3.

Skewness measures the lack of symmetry in a distribution. A positive skewness (between 0.5 and 1) means the tail on the right is longer or fatter whereas a negative skewness (between -1 and -0.5) is when the tail of the left side is longer or fatter.
When the distribution is positively skewed, the mode and median are on the left of the mean whereas when the distribution is negatively skewed, they are on the right.

Kurtosis is used to describe the extreme values in one versus the other tail of the distribution.
A high kurtosis means the data have heavy tails or outliers.
A low kurtosis means the data has light tails or lack of outliers.
A distribution with a kurtosis larger than 3 is called leptokurtic and if less than 3 it is called platykurtic.

Pearson parameters can thus be calculated with skewness (\ref{eq:Skewness}) and kurtosis (\ref{eq:Kurtosis}) by
\begin{align}
c_{0} & =-\frac{\mu_{2}(4\beta_{2}-3\beta_{1}^{2})}{A},\nonumber \\
c_{1} & =~a~=-\frac{\sqrt{\mu_{2}}\beta_{1}(\beta_{2}+3)}{A},\nonumber \\
c_{2} & =-\frac{2\beta_{2}-3\beta_{1}^{2}-6}{A},\nonumber \\
\textrm{where}~A & =10\beta_{2}-12\beta_{1}^{2}-18.
\end{align}

\begin{figure}[H]
\centering \includegraphics[width=10cm]{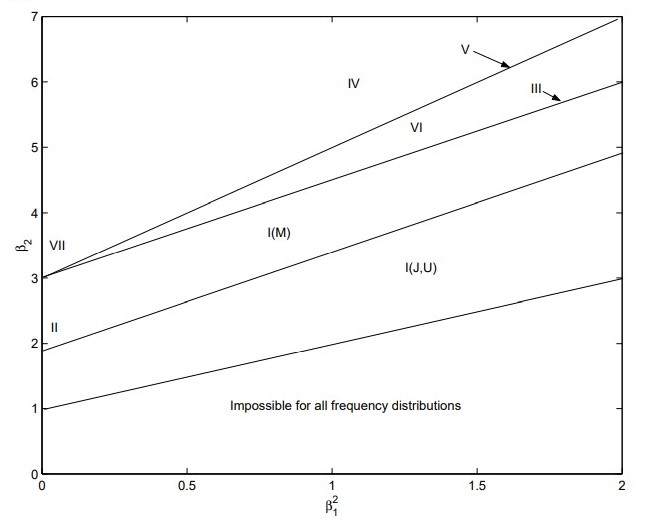}\\
 \caption{Moment ratio diagram for the Pearson curves}
\label{fg:Moment ratio diagram for the Pearson curves}
\end{figure}

The families of Pearson distributions can be specified
in terms of $(\beta_{1}^{2},\beta_{2})$. These can be shown in a
diagram \ref{fg:Moment ratio diagram for the Pearson curves} (screen capture of \cite{4}).

Types of the Pearson distribution are not needed to be specified in this thesis, if they are needed, they can be computed by a program of Qing Yang \& Wei Pan\cite{49}.

The application of the Pearson system to the estimation of the marginal distribution in Tweedie's formula is now evident: given the first four moments we estimate $\frac{g'(x)}{g(x)}$. Table \ref{Algorithm: Pearson system within Tweedie's formula}
shows how the estimation proceeds:
\begin{table}[H]
\centering %
\begin{tabular}{|l|}
\hline
1. Estimate four moments $\mu_{1},\mu_{2},\mu_{3},\mu_{4}$ from data
then compute skewness $\beta_{1}$ and kurtosis $\beta_{2}$.\tabularnewline
2. Calculate the Pearson parameters $a,c_{0},c_{1}$ and $c_{2}$. \tabularnewline
3. Plug in the formula (\ref{eq:Pearson system}) $\frac{g'(x)}{g(x)}=\frac{x-a}{c_{0}+c_{1}x+c_{2}x^{2}}$. \tabularnewline
4. Use Tweedie's formula to obtain the estimation for the posterior
mean or variance. \tabularnewline
\hline
\end{tabular}\caption{Algorithm: Pearson system within Tweedie's formula}
\label{Algorithm: Pearson system within Tweedie's formula}
\end{table}

In this way, the derivative of the logarithm of a density is given as
a rational function of the first four moments of the distribution. The use of the Pearson system has three distinct advantages. First, it eliminates
the need to actually determine the distribution itself. Secondly, it makes use of the four moments so it will be more accurate. It works very well in situations where the variance of a distribution depends on the kurtosis. Finally, the sample size doesn’t need to be very large.

As well, we can take a second derivative:
\begin{align}
[\frac{g'(x)}{g(x)}]'=-\frac{c_{2}x^{2}-2ac_{2}x-(ac_{1}+c_{0})}{(c_{0}+c_{1}x+c_{2}x^{2})^{2}}, \label{eq:Pearson second derivative}
\end{align}
formula (\ref{eq:Pearson second derivative}) is used when we want to compute the posterior variance.

Note that the formula for the derivatives of the log is valid for densities that obey the Pearson system such as normal distribution and gamma distribution. Not every density obeys that system.
For example, the logistic distribution does not satisfy that equation.

Another thing to note is that the Pearson system applies to the unimodal case.
So we need to verify from a histogram of the data to see if it is unimodal or bimodal.
If it is bimodal, we may apply the EM algorithm to compute the weighting coefficient of two distributions obtained by the Pearson system. We do not pursue this subject in this thesis.

\begin{description}
\item[{Application}] Micro-array data
\end{description}
\addcontentsline{toc}{section}{Application: Micro-array data}

To illustrate the application, we recap the following example from Efron.
\footnote{\emph{https://web.stanford.edu/~hastie/CASI/data.html}}
In a study of prostate cancer, n = 102 men each had his genetic expression level $x_{ij}$ measured on N = 6033 genes,
\begin{align}
x_{ij}=\begin{cases}
i=1,2,\cdots,N~genes\\
j=1,2,\cdots,n~men
\end{cases}
\end{align}
There are $n_{1}$ = 50 healthy individuals to serve as controls and $n_{2}$ = 52 prostate cancer patients. The aim is to measure whether case and control are the same for different genes. So for $\textrm{gene}_{i}$
let $t_{i}$ be the two-sample t statistic comparing patients with
controls and
\begin{align*}
z_{i}=\Phi^{-1}[F_{100}(t_{i})];
\end{align*}
$F_{100}$ is cdf of the Student $t_{100}$ distribution with 100
degrees of freedom.

$z_{i}$ is a statistic having a standard normal distribution.
$z_{i}=0$ indicates there is no difference in the case and control groups for $gene_{i}$. (Note: in terms
of our previous notation, $x=z$ .) We recall the Tweedie's formula (\ref{eq:Tweedie normal mean}) in the case of a normal distribution:
\begin{align*}
E[\mu_{i}|z_{i}]=z_{i}+\sigma^{2}(\frac{g_{Z}'(z_{i})}{g_{Z}(z_{i})}).
\end{align*}

Here we assume the variance $\sigma^{2}$ is known to be 1 just as Efron did. If the variance is unknown, we can substitute with the estimator (\ref{eq:Normal variance estimator}) we proposed earlier.

We make use of the Pearson system to estimate the function $\frac{g_{Z}'(\boldsymbol{z})}{g_{Z}(\boldsymbol{z})}$.

Follow algorithm \ref{Algorithm: Pearson system within Tweedie's formula},
we computes the moments to be:
\begin{align*}
\mu_{1}=0.0030,~\mu_{2}=1.2885,~skewness=0.0017,~kurtosis=3.6445.
\end{align*}

So the marginal distribution resembles a standard normal distribution but not completely normal, because kurtosis doesn’t equal to 3.

Thus we obtain $c_{0}=-1.019168,~c_{1}=-0.017116,~c_{2}=-0.069679,a=-0.017116$ and $A=18.42417$, then we can plug $\frac{g'(x)}{g(x)}=\frac{x+0.017116}{-0.069679x^{2}-0.017116x-1.019168}$
in Tweedie's formula.

After the empirical Bayes estimation, we have computed the posterior mean $E[\mu_{i}|z_{i}]$ which we plot against $z_{i}$:
\begin{figure}[H]
\centering \includegraphics[width=12cm]{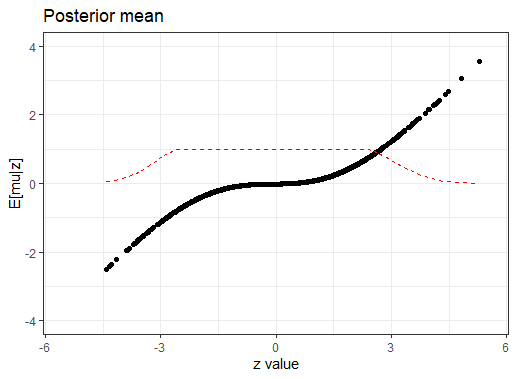}\\
 \caption{Posterior mean for microarray data}
\label{fg:Posterior mean for microarray data}
\end{figure}

Figure \ref{fg:Posterior mean for microarray data} is flatter compared with Efron's plot\cite{26}.
The black points are the estimates which bring the posterior means closer to the overall mean 0. They are near zero ("nullness") when $|z| \leqslant 2$. There are 17 genes flagged (\# of gene: 332 364 579 610 914 1068 1077 1089 1113 1557 1720 3375 3647 3940 4331 4518 4546) that have absolute values larger than 2. At z = 5.29, which is the largest observed $z_{i}$
value (gene \#610), Efron has $E[\mu|z]$ = 3.94, while we have its value is 3.56. So we get smaller values for the outliers.

Dashed curve is estimated local false discovery rate fdr(z)\cite{24}, which is the conditional probability of a case being null given z, declines from one near z=0 to zero at the extremes.
There are 186 genes having fdr(z)$\leqslant$0.2 (a reasonable cutoff point), which might be reported to the researches as candidates for further study.

Then we can plot the posterior variance in the same spirit:
\begin{figure}[H]
\centering \includegraphics[width=12cm]{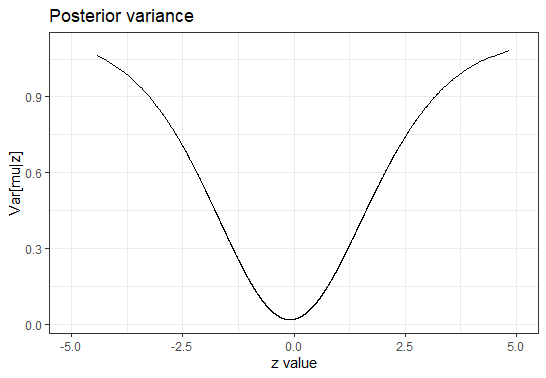}\\
 \caption{Posterior variance for microarray data}
\label{fg:Posterior variance for microarray data}
\end{figure}

The variance is a measure of uncertainty/error. From Figure \ref{fg:Posterior variance for microarray data}
the variance is smallest at 0 and grows on either end as z gets further away from 0. So that's what we expected.

\section{Credible interval}

We can use the posterior mean and variance to provide a credible interval in the Bayesian sense.\cite{23}

In the normal case,
\begin{align}
E(\mu|x)=x+\sigma^{2}\frac{g_{X}'(x)}{g_{X}(x)}~,~~Var({\mu}|x)=\sigma^{2}x+\sigma^{4}[\frac{g_{X}''(x)}{g_{X}(x)}-(\frac{g_{X}''(x)}{g_{X}(x)})^{2}].
\end{align}
where $\frac{g_{X}'(x)}{g_{X}(x)}$ is obtained from the Pearson system.

So the 95\% credible interval for $\mu$ is
\begin{align}
P(a\leqslant\mu\leqslant b|x)=\int_{a}^{b}g(\mu|x)=0.95 \nonumber
\end{align}

If we use a conjugate prior, then we know the posterior distribution
is also normal. The credible interval is given by
\begin{align}
(qnorm(0.025,E({\mu}|x),Var({\mu}|x)),qnorm(0.975,E({\mu}|x),Var({\mu}|x)))
\end{align}

In the case of Poisson distribution, $E[log\theta|x]=\frac{g_{X}'(x)}{g_{X}(x)}-\frac{g_{0}'(x)}{g_{0}(x)}$,
so
\begin{align}
E[\theta|x]=e^{\frac{g_{X}'(x)}{g_{X}(x)}-\frac{g_{0}'(x)}{g_{0}(x)}}.
\end{align}

In the case of Binomial, we computed the posterior odds $E[log\frac{\theta}{1-\theta}|x]=\frac{g_{X}'(x)}{g_{X}(x)}-\frac{g_{0}'(x)}{g_{0}(x)}$,
so
\begin{align}
E[p|x]=\frac{e^{\frac{g_{X}'(x)}{g_{X}(x)}-\frac{g_{0}'(x)}{g_{0}(x)}}}{1+e^{\frac{g_{X}'(x)}{g_{X}(x)}-\frac{g_{0}'(x)}{g_{0}(x)}}}.
\end{align}

\section{Saddlepoint generalization}

H. Daniels (1954)\cite{15} proposed the saddlepoint method in order to
obtain a highly accurate approximation formula for any probability
density function or probability mass function of a distribution, based
on the moment generating function. Here, we briefly describe the method
and then apply it in our context.

\subsection{Derivation}

Suppose we have a density from the exponential family defined as:
\begin{align}
f_{X}(x;\theta) & =exp\{\theta x-K(\theta)-d(x)\}.
\end{align}

We shall sketch the derivation of the saddlepoint approximation from Glen\_b\cite{29}. The above density is given by
\begin{align}
f_{X}(x;\theta)\approx(\frac{1}{2\pi K''(\hat{\theta})})^{\frac{1}{2}}exp\big\{ K(\hat{\theta})-\hat{\theta}x\big\},
\end{align}
where $K(\cdot)$ is the cumulant generating function of the given distribution
and $\hat{\theta}$ is the maximum likelihood estimate of $\theta$.\\

We begin with the assumption that the moment generating function exists
and is twice differentiable. This implies in particular that all moments
exist. Let X be a random variable with moment generating function
$M(t)=e^{K(t)}$. Consider the Laplace approximation to the following integral.
\begin{align}
e^{K(t)}=\int_{-\infty}^{\infty}e^{tx}f(x)dx=\int_{-\infty}^{\infty}exp(tx+\log f(x))dx=\int_{-\infty}^{\infty}exp(-h(t,x))dx,
\end{align}
where
\begin{align*}
h(t,x)=-tx-\log f(x).
\end{align*}

Expand $h(t,x)$ in a Taylor series around $x_{0}$ keeping t constant
to obtain
\begin{align}
h(t,x)=h(t,x_{0})+h'(t,x_{0})(x-x_{0})+\frac{1}{2}h''(t,x_{0})(x-x_{0})^{2}+\cdots,
\end{align}
where ' denotes differentiation with respect to x. Letting $x_{t}$
be the solution to $h'(t,x_{t})=0$ yields the minimum for $h(t,x)$
as a function of x.
\begin{align*}
e^{K(t)}\approx\int_{-\infty}^{\infty}exp(-h(t,x_{t})-\frac{1}{2}h''(t,x_{t})(x-x_{t})^{2})dx=e^{-h(t,xt)}\int_{-\infty}^{\infty}e^{-\frac{1}{2}h''(t,x_{t})(x-x_{t})^{2}}dx,
\end{align*}
which is a Gaussian integral, giving:
\begin{align}
e^{K(t)}\approx e^{-h(t,x_{t})}\sqrt{\frac{2\pi}{h''(t,x_{t})}}.
\end{align}

This can then be transformed into the saddlepoint approximation:
\begin{align}
e^{K(t)} & \approx e^{tx_{t}+\log f(x_{t})}\sqrt{\frac{2\pi}{h''(t,x_{t})}}, \nonumber \\
f(x_{t}) & \approx\sqrt{\frac{h''(t,x_{t})}{2\pi}}exp[K(t)-tx_{t})].
\end{align}

From $h'(t,x_{t})=-t-\frac{\partial \log f(x_{t})}{\partial x_{t}}=0$ we get:
\begin{align}
K'(t)=x_{t} \label{eq:Saddlepoint equation}
\end{align}

Taking second derivative yields:
\begin{align}
h''(t,x_{t}) & =-\frac{\partial^{2}\log f(x_{t})}{\partial^{2}x_{t}}=-\frac{\partial}{\partial x_{t}}(-t)=\frac{\partial x_{t}}{\partial t} \nonumber\\
 & =\frac{1}{K'(t)}.
\end{align}

Hence, to approximate the density at a specific point $x_{t}$, we
solve the saddlepoint equation for that $x_{t}$ to find t. The final
expression for the saddlepoint approximation of the density $f(x)$
is given by
\begin{align*}
f(x_{t})\approx\sqrt{\frac{1}{2\pi K''(t)}}e^{K(t)-tx_{t}}.
\end{align*}

\subsection{Accuracy of saddlepoint approximation}

It is of interest to check on the accuracy of the saddlepoint approximation in various examples of distributions.
(Details of derivations are included in Appendix A.1.2).
\begin{description}
\item [{Example 1}] Normal distribution
\end{description}

The exponential form of the Normal density is given by
\begin{align*}
f(x|\mu) & =\frac{1}{\sqrt{2\pi\sigma_{0}^{2}}}e^{\frac{(x-\mu)^{2}}{2\sigma_{0}^{2}}},\\
f(x|\eta) & =exp\{\eta x-\frac{\sigma_{0}^{2}\eta^{2}}{2}\}\frac{1}{\sqrt{2\pi\sigma_{0}^{2}}}e^{-\frac{x^{2}}{2\sigma_{0}^{2}}}.
\end{align*}

The saddlepoint approximation gives
\begin{align}
f(x=\hat{\eta};\eta) & \approx(2\pi)^{-\frac{1}{2}}(\sigma_{0}^{2})^{-\frac{1}{2}}exp\big\{(\eta-\frac{x}{\sigma_{0}^{2}})x-(\frac{\sigma_{0}^{2}\eta^{2}}{2}-\frac{x}{2\sigma_{0}^{2}})\big\}
\nonumber \\
 & \approx(2\pi\sigma_{0}^{2})^{-\frac{1}{2}}exp\big\{(\eta x-\frac{\sigma_{0}^{2}\eta^{2}}{2}-\frac{x^{2}}{2\sigma_{0}^{2}})\big\}.
\end{align}

The result in this case is exact.

\begin{description}
\item [{Example 2}] Exponential distribution
\end{description}

The exponential form of the Exponential distribution is given by
\begin{align*}
f(x|\theta) & =\theta e^{-\theta x}\\
 & =exp\{\theta(-x)+log(\theta)\}.
\end{align*}

The saddlepoint approximation gives
\begin{align}
f(x;\theta) & \approx(2\pi)^{-\frac{1}{2}}(x^{2})^{-\frac{1}{2}}exp\big\{(\theta-\frac{1}{x})(-x)-(-log\theta+log\frac{1}{x})\big\} \nonumber \\
 & \approx(2\pi)^{-\frac{1}{2}}x^{-1}exp\big\{(-\theta x+1)+log\theta-log\frac{1}{x})\big\} \nonumber \\
 & \approx\frac{e}{\sqrt{2\pi}}exp\big\{\theta(-x)+log\theta)\big\}.
\end{align}

We calculate the ratio
\[
\frac{e}{\sqrt{2\pi}}=2.718/2.5066=1.08.
\]
So this approximation is roughly exact.

\begin{description}
\item [{Example 3}] Poisson distribution
\end{description}

The exponential form of Poisson distribution is given by
\begin{align*}
f(x|\theta) & =\frac{\theta^{x}}{x!}e^{-\theta}\\
 & =exp\{log\theta \cdot x-\theta-logx!\},\\
f(x|\theta)  & =exp\{\eta x-e^{\eta}-logx!\}.
\end{align*}

The saddlepoint approximation gives
\begin{align}
f(x;\eta) & \approx(2\pi)^{-\frac{1}{2}}(x)^{-\frac{1}{2}}exp\big\{(\eta-logx)(x)-(e^{\eta}-x)\big\} \nonumber \\
 & \approx(2\pi x)^{-\frac{1}{2}}exp\big\{\eta x-e^{\eta}-logxx+x)\big\} \nonumber \\
 & \approx exp\big\{\eta x-e^{\eta}\big\}\sqrt{\frac{1}{2\pi x}}(\frac{e}{x})^{x}.
\end{align}

Consequently, $\sqrt{\frac{1}{2\pi x}}(\frac{e}{x})^{x}$ approximates
$\frac{1}{x!}$. which is actually the Stirling formula for factorials.

\begin{description}
\item [{Example 4}] Binomial distribution
\end{description}

The exponential form of Binomial distribution is given by
\begin{align*}
f(x|p) & =\binom{n}{x}p^{x}(1-p)^{n-x}\\
 & =exp\{\log\frac{p}{1-p}x+n\log1-p\}\binom{n}{x},\\
f(x|\eta) & =exp\{\eta x+n\log\frac{1}{e^{\eta}+1}\}\binom{n}{x}.
\end{align*}

The saddlepoint approximation gives
\begin{align}
f(x;\eta) & \approx(2\pi)^{-\frac{1}{2}}(\frac{(n-x)x}{n})^{-\frac{1}{2}}exp\big\{(\eta-log\frac{x}{n-x})x-(n\log(e^{\eta}+1)-n\log(\frac{n}{n-x})\big\} \nonumber \\
 & \approx(2\pi\frac{(n-x)x}{n})^{-\frac{1}{2}}exp\big\{(\eta x-n\log(e^{\eta}+1)\big\}(\frac{x}{n-x})^{-x}(\frac{n}{n-x})^{n} \nonumber \\
 & \approx exp\big\{(\eta x-n\log(e^{\eta}+1)\big\}(\frac{x}{n-x})^{-x}(\frac{n}{n-x})^{n}\sqrt{\frac{n}{2\pi(n-x)x}}.
\end{align}

Consequently, $(\frac{x}{n-x})^{-x}(\frac{n}{n-x})^{n}\sqrt{\frac{n}{2\pi(n-x)x}}$
is used to approximate $\frac{n!}{x!(n-x)!}$ which once again is Stirling's formula.

\begin{description}
\item [{Example 5}] Gamma distribution
\end{description}

Suppose the shape parameter $\alpha$ is known in the Gamma distribution, its density is given by
\begin{align*}
f(x|\beta) & =\frac{\beta^{\alpha}}{\varGamma\left(\alpha\right)}x^{\alpha-1}e^{-\beta x}\\
 & =exp\{\beta(-x)+\alpha\log\beta\}\frac{x^{\alpha-1}}{\varGamma(\alpha)}.
\end{align*}

The saddlepoint approximation gives
\begin{align}
f(x;\theta) & \approx(2\pi)^{-\frac{1}{2}}(\frac{x^{2}}{\alpha})^{-\frac{1}{2}}exp\big\{(\beta-\frac{\alpha}{x})(-x)-(-\alpha\log\beta+\alpha\log\frac{\alpha}{x})\big\}
\nonumber\\
 & \approx(\frac{\alpha}{2\pi x^{2}})^{\frac{1}{2}}exp\big\{-\beta x+\alpha+\alpha\log\beta+\alpha\log\frac{x}{\alpha}\big\} \nonumber \\
 & \approx exp\big\{-\beta x+\alpha\log\beta\big\}\sqrt{\frac{\alpha}{2\pi x^{2}}}e^{\alpha}(\frac{x}{\alpha})^{\alpha}.
\end{align}

Consequently, $\sqrt{\frac{\alpha}{2\pi}}(\frac{e}{\alpha})^{\alpha}$
serves to approximate $\frac{1}{\varGamma(\alpha)}$. If $\alpha$=1,
this is consistent with exponential distribution.\\

From above examples it can be seen that the Saddlepoint approximation works well on exponential families.
\subsection{Combination with Tweedie's formula}

We now originally make use of the saddlepoint approximation to generalize
Tweedie's formula as the result holds for arbitrary exponential families
and can work for sufficient statistics for $\theta$.

Suppose $X_{1},...,X_{n}$ is a random sample with a density from an exponential family
\begin{align}
g(x;\theta)=exp\{\theta^{T}a(x)-\psi(\theta)-d(x)\},
\end{align}
where the vector $a(x)$ is the minimal sufficient statistic which represents
the data, $\psi(\cdot)$ is the cumulant generating function and $d(\cdot)$ is carrier measure. We will be concerned with the non-null situation where $\theta\neq0$.

The density of $a(x)$ at $a(x)=a$ is given by
\begin{align}
g(a;\theta)=exp\{\theta^{T}a-n\psi(\theta)-d^{*}(a)\}. \label{eq:Density of minimal sufficient statistic}
\end{align}

We may approximate $exp \{-d^{*}(a)\}$ using the saddlepoint approximation\cite{30}. Specifically, we have
\begin{align}
exp\{-d^{*}(a)\} & =g(a;\theta)exp\{\theta^{T}a-n\psi(\theta)\} \nonumber \\
 & \sim(2\pi)^{\frac{1}{2}}(n\psi''(\hat{\theta}))^{-\frac{1}{2}}exp[n\psi(\hat{\theta})-\hat{\theta}^{T}a], \label{eq:Approxiamtion of carrier measure}
\end{align}
where the saddlepoint equation is $n\psi'(\theta)=a$.
The saddlepoint occurs at the maximum likelihood estimate $\hat{\theta}$.

Plug equation (\ref{eq:Approxiamtion of carrier measure}) into equation (\ref{eq:Density of minimal sufficient statistic}), we have the density of $a(x)$ at value $a$
\begin{align}
g(a;\theta)\sim(2\pi)^{\frac{1}{2}}(n\psi''(\hat{\theta}))^{-\frac{1}{2}}exp\big\{(\theta-\hat{\theta})^{T}a-n(\psi(\theta)-\psi(\hat{\theta}))\big\}.
\end{align}

The marginal distribution is given by
\begin{align*}
g_{X}(a) & =\int g(a;\theta)\pi(\theta)d\theta\\
 & \sim\int(2\pi)^{\frac{k}{2}}(n\psi''(\hat{\theta}))^{-\frac{1}{2}}exp\big\{(\theta-\hat{\theta})^{T}a-n(\psi(\theta)-\psi(\hat{\theta}))\big\}\pi(\theta)d\theta.
\end{align*}

Suppose $n=1,a=x$, then $\psi'(\hat{\theta})=x$.

We can view $\hat{\theta}$ is a function of x, compute derivatives of the marginal:
\begin{align*}
g_{X}'(x) & =\int[(2\pi)^{\frac{1}{2}}(\psi''(\hat{\theta}))^{-\frac{1}{2}}exp\big\{(\theta-\hat{\theta})^{T}a-(\psi(\theta)-\psi(\hat{\theta}))\big\}]'\pi(\theta)d\theta\\
 & =[(\psi''(\hat{\theta}))^{-\frac{1}{2}}]'\int(2\pi)^{\frac{1}{2}}exp\big\{(\theta-\hat{\theta})^{T}a-(\psi(\theta)-\psi(\hat{\theta}))\big\}\pi(\theta)d\theta\\
 & ~~+\int[(\theta-\hat{\theta})^{T}a-(\psi(\theta)-\psi(\hat{\theta})]'(2\pi)^{\frac{1}{2}}(\psi''(\hat{\theta}))^{-\frac{1}{2}}exp\big\{(\theta-\hat{\theta})^{T}a-(\psi(\theta)-\psi(\hat{\theta}))\big\}\pi(\theta)d\theta\\
 & =-\frac{1}{2}\frac{\psi'''(\hat{\theta})}{[\psi''(\hat{\theta})]}\frac{d}{dx}\hat{\theta}g(x)+E[\theta-\hat{\theta}|x]g(x)-x\frac{d}{dx}\hat{\theta}g(x)+\psi'(\hat{\theta})\frac{d}{dx}\hat{\theta}g(x),
\end{align*}
which would give us
\begin{align}
\frac{g_{X}'(x)}{g_{X}(x)} & =E[\theta-\hat{\theta}|x]-x\frac{d}{dx}\hat{\theta}+\psi'(\hat{\theta})\frac{d}{dx}\hat{\theta}-\frac{1}{2}\frac{\psi'''(\hat{\theta})}{[\psi''(\hat{\theta})]}\frac{d}{dx}\hat{\theta} \nonumber \\
 & =E[\theta-\hat{\theta}|x]-\frac{1}{2}\frac{\psi'''(\hat{\theta})}{[\psi''(\hat{\theta})]}\frac{d}{dx}\hat{\theta}.
\end{align}

So we derived Tweedie's formula in the general case of an exponential family:
\begin{align}
E[\theta|x]=h(x)+\frac{1}{2}\frac{\psi'''(\hat{\theta})}{[\psi''(\hat{\theta})]}\frac{d}{dx}\hat{\theta}+\hat{\theta},
\end{align}
where $h(x)=\frac{g_{X}'(x)}{g_{X}(x)}$.

In general we have properties of $\frac{g_{X}'(x)}{g_{X}(x)}$ from
the Pearson system of distributions, but we do not have an explicit
formula for
\begin{align*}
\hat{\theta}+\frac{1}{2}\frac{\psi'''(\hat{\theta})}{[\psi''(\hat{\theta})]}\frac{d}{dx}\hat{\theta}
\end{align*}

So if we have the form of the cumulant function, we can get the Tweedie's estimates for the posterior means for $\theta$.

From the Table \ref{tb:Computations of some exponential family distributions} below, $\hat{\theta}+\frac{1}{2}\frac{\psi'''(\hat{\theta})}{[\psi''(\hat{\theta})]}\frac{d}{dx}\hat{\theta}$
has the form $\hat{\theta}+\frac{c}{x}$, where c is a constant,
and the second term has little influence compared with the first term.
So we can approximate the estimator with $\hat{\theta}$ which can be obtained from maximum likelihood estimation.

\begin{table}[H]
\resizebox{\textwidth}{!}{ %
\begin{tabular}{|c|c|c|c|c|c|}
\hline
Density  & $f\left(x\right)$  & $\psi\left(t\right)$  & $\psi'\left(t\right)$  & $\frac{\psi'''\left(t\right)}{\psi''\left(t\right)}$  & $\hat{\theta}+\frac{1}{2}\frac{\psi'''\left(\hat{\theta}\right)}{\left[\psi''\left(\hat{\theta}\right)\right]}\frac{d}{dx}\hat{\theta}$\tabularnewline
\hline
\hline
Normal  & $N\left(0,\sigma^{2}\right)$  & $\frac{\sigma^{2}t^{2}}{2}$  & $\sigma^{2}t$  & $0$  & $\frac{x}{\sigma^{2}}$\tabularnewline
\hline
Laplace{*}  & $\frac{1}{2b}exp\left(-\frac{\left|x-\mu\right|}{b}\right)$  & $\mu t-log\left(1-b^{2}t^{2}\right)$  & $\mu+\frac{2b^{2}t}{1-b^{2}t^{2}}$  & $\frac{2b^{2}\left(3t+b^{2}t^{3}\right)}{1-b^{4}t^{4}}$  & \tabularnewline
\hline
Gamma  & $\frac{\beta^{\alpha}}{\varGamma\left(\alpha\right)}x^{\alpha-1}e^{-\beta x}$  & $-\alpha log\left(1-\frac{t}{\beta}\right)$  & $\frac{\alpha}{\beta-t}$  & $\frac{2}{\beta-t}$  & $\beta-\frac{\alpha}{x}+\frac{1}{x}$\tabularnewline
\hline
Chi square  & $\frac{1}{2^{k/2}\varGamma\left(k/2\right)}x^{k/2-1}e^{-x/2}$  & $-\frac{k}{2}log\left(1-2t\right)$  & $\frac{k}{1-2t}$  & $\frac{4}{1-2t}$  & $\frac{1}{2}-\frac{k}{2x}+\frac{1}{x}$\tabularnewline
\hline
Exponential  & $\lambda e^{-\lambda x}$  & $log\frac{\lambda}{\lambda-t}$  & $\frac{1}{\lambda-t}$  & $\frac{2}{\lambda-t}$  & $\lambda-\frac{\lambda}{e^{x}}+\frac{\lambda}{x}$\tabularnewline
\hline
Beta  & $\frac{\varGamma(\alpha)\varGamma(\beta)}{\varGamma(\alpha+\beta)}x^{\alpha-1}(1-x)^{\beta-1}$  & $log(1+\overset{\infty}{\underset{k=1}{\sum}}(\overset{k-1}{\underset{r=0}{\prod}}\frac{\alpha+\beta}{\alpha+\beta+r})\frac{t^{k}}{k})$  &  &  & \tabularnewline
\hline
Possion  & $\frac{\lambda^{x}e^{-\lambda}}{k!}$  & $\lambda(e^{t}-1)$  & $\lambda e^{t}$  & 1  & $log(\frac{x}{\lambda})+\frac{1}{2x}$\tabularnewline
\hline
Binomial  & $\binom{n}{x}p^{x}(1-p)^{n-x}$  & $nlog(1-p+pe^{t})$  &  &  & \tabularnewline
\hline
Geometric  & $(1-p)^{x-1}p$  & $log(\frac{pe^{t}}{1-(1-p)e^{t}})$  &  &  & \tabularnewline
\hline
\end{tabular}} {*} not a member of the exponential family of distributions \caption{Computations of some exponential family distributions}
\label{tb:Computations of some exponential family distributions}
\end{table}

\section{Application to ranking data}

Our contribution in this section is to apply Tweedie's formula in the context of rankings and test its effects on real data sets.

Suppose that $\mathbf{R}$ represents a ranking of $t$ objects, it is convenient to standardize the
rankings under the assumption of uniformity by subtracting the null mean $\frac{t+1}{2}$ and dividing by the standard deviation $\sqrt{\frac{t(t^{2}-1)}{12}}$:
\begin{align}
\boldsymbol{x}=\frac{\boldsymbol{R}-\frac{t+1}{2}}{\sqrt{\frac{t(t^{2}-1)}{12}}}.
\end{align}

Then its density is given by
\begin{align}
g_{\boldsymbol{\theta}}(\boldsymbol{x})=exp\{\boldsymbol{\theta}^{T}\boldsymbol{x}-K(\boldsymbol{\theta})\}g_{0}(\boldsymbol{x}).
\end{align}

The density $g_{0}(x)$ can take different forms at our
disposal and it represents the null situation when $\boldsymbol{\theta}=\mathbf{0}.$

Let $\pi(\boldsymbol{\theta})$ be a prior density on $\boldsymbol{\theta}$. Then
the marginal density of $X$ is
\begin{align}
g_{X}(\boldsymbol{x}) & = \int g(\boldsymbol{x};\boldsymbol{\theta})\pi(\boldsymbol{\theta})d\boldsymbol{\theta} \nonumber \\
 & = \int exp\{ \boldsymbol{\theta}^{T}\boldsymbol{x}-K(\phi)\} g_{0}(\boldsymbol{x})\pi(\boldsymbol{\theta})d\boldsymbol{\theta}.
\end{align}

Now taking derivative
\begin{align*}
\frac{\partial}{\partial \boldsymbol{x}}g_{X}(\boldsymbol{x}) & =\int\theta exp\{\theta^{T}\boldsymbol{x}-K(\phi)\}g_{0}(\boldsymbol{x})\pi(\boldsymbol{\theta})d\boldsymbol{\theta}+\frac{\partial}{\partial \boldsymbol{x}}g_{0}(\boldsymbol{x})\int exp\{\boldsymbol{\theta}^{T}\boldsymbol{x}-Kt(\phi)\}\pi(\boldsymbol{\theta})d\boldsymbol{\theta},\\
\frac{1}{g_{X}(\boldsymbol{x})}\frac{\partial}{\partial \boldsymbol{x}}g_{X}(\boldsymbol{x}) & =E[\boldsymbol{\theta}|\boldsymbol{x}]+\frac{1}{g_{0}(\boldsymbol{x})}\frac{\partial}{\partial \boldsymbol{x}}g_{0}(\boldsymbol{x}),
\end{align*}

It follows that (applying Tweedie's formula component by component)
\begin{equation}
E[\boldsymbol{\theta}|\boldsymbol{x}]=\frac{1}{g_{X}(\boldsymbol{x})}\frac{\partial}{\partial \boldsymbol{x}}g_{X}(\boldsymbol{x})-\frac{1}{g_{0}(x)}\frac{\partial}{\partial \boldsymbol{x}}g_{0}(\boldsymbol{x}).\label{eq:Tweedie's formula on ranking}
\end{equation}

The first term at the right hand side of equation (\ref{eq:Tweedie's formula on ranking})
is a vector whose components are $(\frac{\partial}{\partial x_{i}}log\:g_{X}(x_{i}))$.
We will use the Pearson system on these components, one at a time and estimate them using each column of the data $\{\boldsymbol{x}\}$.
We shall consider the carrying density $g_{0}(\boldsymbol{x})$.

\subsection{Three situations}

In this section, we consider three examples of densities
$g_{0}(\boldsymbol{x})$. We then apply the result to detecting differences
in Sushi food preference patterns.

\subsubsection{Uniform case}

Here, we assume that the rankings of $t$ objects are equally likely so that
\begin{align*}
g_{0}(\boldsymbol{x})=\frac{1}{t!},~\textrm{for all}~\boldsymbol{x}
\end{align*}

It follows that
\begin{align*}
\frac{1}{g_{0}(\boldsymbol{x})}\frac{\partial}{\partial\boldsymbol{x}}g_{0}(\boldsymbol{x})=0,
\end{align*}

and consequently
\begin{align}
E[\boldsymbol{\theta}|\boldsymbol{x}]=\frac{1}{g_{X}(\boldsymbol{x})}\frac{\partial}{\partial\boldsymbol{x}}g_{X}(\boldsymbol{x}).
\end{align}

\subsubsection{Von-Mises distribution}

Consider the following angle-based model proposed by M. Alvo \& Hang Xu\cite{48}, where they assumed a consensus score vector $\boldsymbol{m}$ and the probability of observing a ranking $\boldsymbol{x}$ is proportional to the cosine of the angle from $\boldsymbol{m}$.
Our carrying density $g_{0}(x)$ will be written as
\begin{align}
g_{0}(\boldsymbol{x}) & =C(\kappa,\boldsymbol{m})exp\{\kappa\boldsymbol{m}^{T}\boldsymbol{x}\}, \label{eq:Angle-based model}
\end{align}
where the parameter $\Vert\boldsymbol{m}\Vert=1=\Vert\boldsymbol{x}\Vert$, parameter
$\kappa\geqslant0$, and $C(\kappa,\boldsymbol{m})$ is the normalizing constant.

To compute the normalizing constant $C(\kappa,\boldsymbol{m})$, let
$P_{t}$ be the set of all possible permutations of the integers 1,...,t. Then
\begin{align}
(C(\kappa,\boldsymbol{m}))^{-1}=\underset{x\in P_{t}}{\sum}exp\{\kappa\boldsymbol{m}^{T}\boldsymbol{x}\}. \label{eq:Sum of normalizing constant}
\end{align}

When t is large (larger than 15), the exact calculation of the normalizing constant is difficult to compute.
This is where the von Mises distribution comes in. It can be used to work out an exact expression for the constant. The von Mises distribution is for continuous data on a sphere. The ranking data is discrete and it can also be thought of as placed on a t!-dimensional sphere, so it resembles this distribution. Some empirical evidence showed that for a number of different values of t the approximation works fairly well.\cite{48}
The continuous von Mises-Fisher distribution, abbreviated as $vMF(\boldsymbol{x}|\boldsymbol{m},\kappa)$, its density is defined as:
\begin{align}
p(\boldsymbol{x}|\kappa,\boldsymbol{m})=V_{t}(\kappa)exp(\kappa\boldsymbol{m}^{T}\boldsymbol{x}), \label{eq:von Mises distribution}
\end{align}
where $V_{t}(\kappa)=\frac{\kappa^{\frac{t}{2}-1}}{(2\pi)^{\frac{t}{2}}I_{\frac{t}{2}-1}(\kappa)}$ and
$I_{\frac{t}{2}-1}$ is the modified Bessel function of the first kind with order $\frac{p}{2}-1$.

Consequently, the sum (\ref{eq:Sum of normalizing constant}) can be approximated by an integral over the sphere
\begin{align*}
C(\kappa,\boldsymbol{m})\approx C_{t}(\kappa)=\frac{\kappa^{\frac{t-3}{2}}}{2^{\frac{t-3}{2}}t!I_{\frac{t-3}{2}}(\kappa)\Gamma(\frac{t-1}{2})}.
\end{align*}

Then
\begin{align}
\frac{\partial}{\partial\boldsymbol{x}}g_{0}(\boldsymbol{x})&=\kappa\boldsymbol{m}\cdot C_{t}(\kappa) exp\{\kappa\boldsymbol{m}^{T}\boldsymbol{x}\}, \nonumber \\
\frac{1}{g_{0}(\boldsymbol{x})}\frac{\partial}{\partial\boldsymbol{x}}g_{0}(\boldsymbol{x})&=\kappa\boldsymbol{m}.
\label{eq:Derivatives of von Mises case}
\end{align}

By maximum likelihood estimation we get:
\begin{align*}
\hat{\boldsymbol{m}} & =\frac{\sum_{i=1}^{N}\boldsymbol{x}_{i}}{\|\sum_{i=1}^{N}\boldsymbol{x}_{i}\|},\\
\hat{\kappa} & =\frac{r(t-1-r^{2})}{1-r^{2}}.\\
\textrm{where}~r & =\frac{\|\sum_{i=1}^{N}\boldsymbol{x}_{i}\|}{N}
\end{align*}

Plug equation (\ref{eq:Derivatives of von Mises case}) in equation (\ref{eq:Tweedie's formula on ranking}), the posterior mean will then be in the form
\begin{align}
E(\boldsymbol{\theta}|\boldsymbol{x})=\frac{1}{g_{X}(\boldsymbol{x})}\frac{\partial}{\partial\boldsymbol{x}}g_{X}(\boldsymbol{x})-\frac{r(t-1-r^{2})}{1-r^{2}}\frac{\sum_{i=1}^{N}\boldsymbol{x}_{i}}{\|\sum_{i=1}^{N}\boldsymbol{x}_{i}\|}.
\end{align}

\subsubsection{Multivariate Normal distribution}

The normal distribution is frequently used in practice to model different phenomena. In the present context we use it to model ranking data. Assume the conditional probability distribution of $\mathbf{x}$, $g_{\boldsymbol{\theta}}(\boldsymbol{x}),$
is the multivariate normal distribution $N(\boldsymbol{\mu},\varSigma)$. This indicates the carrying density $g_{0}(\boldsymbol{x})$ is the multivariate standard normal distribution.

Using the results (\ref{eq:Posterior mean of multivariate normal}) we obtained in Section 3.3.1, we see immediately that the posterior mean is
\begin{align}
E(\boldsymbol{\mu}|\boldsymbol{x})=\varSigma\frac{1}{g(\boldsymbol{x})}\frac{\partial}{\partial \boldsymbol{x}}g(\boldsymbol{x})+\boldsymbol{x}.
\end{align}

\begin{description}
\item [{Application}] Sushi data
\end{description}
\addcontentsline{toc}{section}{Application: Sushi data}

In this application, we make use of the sushi data described in Kamishima (2003)\cite{36}.
Our interest is to uncover differences in food preference patterns between the populations in eastern and western
Japan.
Historically, western Japan has been mainly affected by the culture of the Mikado emperor and nobles, while eastern Japan has been the home of the Shogun and Samurai warriors. Therefore, the preferences in sushi food might be quite different between these two regions.\cite{36}

Here we view eastern and western people are from different distributions. We assume each individual has a ranking (preferences of t=10 different kinds of sushi) and an underlying $\theta_{i}$. We computed the posterior $\theta$'s for all the individuals and then calculated the mean for the eastern and western populations. We rank the posterior mean for the preference and display in Table \ref{tb:Results of sushi preference rankings of Eastern and Western Japanese}.
\begin{table}[H]
\resizebox{\textwidth}{!}{ \centering %
\begin{tabular}{ccccccccccc}
\hline
 & shrimp  & sea eel  & tuna  & squid  & sea urchin  & salmon roe  & egg  & fatty tuna  & tuna roll  & cucumber\tabularnewline
$\textrm{Eastern}_{\textrm{von-Mises}}$:  & 5  & 6  & 2  & 8  & 4  & 3  & 9  & 1  & 7  & 10\tabularnewline
$\textrm{Western}_{\textrm{von-Mises}}$:  & 6  & 5  & 2  & 7  & 3  & 4  & 9  & 1  & 8  & 10\tabularnewline
$\textrm{Eastern}_{\textrm{normal}}$:  & 3  & 4  & 5  & 6  & 9  & 1  & 8  & 2  & 7  & 10\tabularnewline
$\textrm{Western}_{\textrm{normal}}$:  & 2  & 1  & 10  & 3  & 5  & 9  & 4  & 6  & 8  & 7\tabularnewline
\hline
\end{tabular}} \caption{Results of sushi preference rankings of Eastern and Western Japanese}
\label{tb:Results of sushi preference rankings of Eastern and Western Japanese}
\end{table}

As you can see, von-Mises case have similar ranking of eastern and western people so it did not detect the difference of preference.
However, the normal case indicated that people in eastern Japan have a greater preference for Salmon roe and fatty Tuna than western people. On the other hand, the latter prefer Sea eel, Shrimp and Squid. This result is in accord with the original research.
This result is in accord with conclusions reached in the original research.

\chapter{Disscusion}

\section{Conclusion}

This study has revisited some empirical Bayes methods and explored their applications.
We start with a simple linear empirical Bayes method where the linear estimator is obtained by minimizing the mean square error. We directly applied it to ranking data sets and obtained consistent predictions with the official sites. Then results are also good on other data sets not displayed in this thesis, such as the UK University Ranking data.\footnote{https://www.thecompleteuniversityguide.co.uk/league-tables/rankings} We find that several un-related groups can contribute to the estimation in each group, more information always provides more accurate estimates.
We also applied it to Interval data while we made some changes to suit the data type.
The estimates can amplify internal characteristics and move towards the overall mean.

We then looked at Tweedie's formula. We proposed the Pearson system to compute the derivative of the log marginal distribution, to obviate the need to specify the entire marginal distribution. And it still works well when the sample size is small (say 15).
Next we generalized the Tweedie's formula using Saddlepoint approximation. The applied results on ranking data are pretty good showing that the methodology can accurately identify the preferences of different groups.

Although this thesis mainly applies empirical Bayes methods on ranking and interval data, I believe they have great application prospects to other types of data.

\section{Future research}

We can further our exploration on empirical Bayesian analysis. I propose two directions so far.

\subsection{Dynamic approach}

In our above setting, the $\theta_{i}$ are assumed to have unknown prior distributions.
Let $d_{G}(x)$ denote the Bayes decision rule when $X_{i}=x$ is observed. The basic principle underlying empirical Bayes is that $d_{G}$ can often be consistently estimated from the data \{$X_{1},...,X_{n}$\},
leading to the empirical Bayes rule $d_{\hat{G}}$. Thus, the n structurally similar problems can be pooled to provide information about unspecified prior distribution, thereby yielding $\hat{G}$ and the decision rules $d_{\hat{G}}(X_{i})$) for the independent problems.\cite{38}

Our next interest is to look at longitudinal data, we need to exploit information from different individuals or objects over different time periods.
Consider the general linear model described by Lai \& Su (2014)\cite{38},
in a general framework, $X_{i,t}$ belongs to an exponential family of distributions, when the ith subject is observed for time t,
the conditional density of $X_{t}$ given
$X_{t-1}$ is given by
\begin{align}
g(x;\theta_{i,t},\phi)=exp\{\frac{x\theta_{i,t}-f(\theta_{i.t})}{\phi}+c(x,\phi)\},
\end{align}
which with the canonical link yields, (if $\phi=1$)
\begin{align}
E\left[X_{i,t}|X_{i,t-1}\right]
 & =  \theta_{i,t} \nonumber \\
 & =  \beta+\rho X_{i,t-1}.
\end{align}

Thus we need to think about how to estimate the parameters $(\beta,\rho,\phi)$. They also propose a model with covariates which includes data for several years. This will further complicate the problem of estimation.

\subsection{Parallel Randomized Experiments}

Nowadays many studies comparing new treatments to standard treatments is composed of parallel randomized experiments. It would be of interest to use empirical Bayes methods to summarize the evidence in data about differences among parallel experiments, thereby obtaining improved estimates for the treatment effect in each experiment from all data combined.\cite{47}

\newpage
\appendix

\chapter{Supplementary materials}
\section{Derivations}
\subsection{Transformation from interval bounds to center and half-width}
\paragraph{Sample mean}
\begin{align*}
\bar{X}& =\frac{1}{n}{\underset{w\in E}{\sum}}\frac{l_{w}+u_{w}}{2}\\
& =\frac{1}{n}{\underset{w\in E}{\sum}}c_{w}.
\end{align*}
\paragraph{Sample variance}
\begin{align*}
S^{2} & =\frac{1}{3n}{\underset{w\in E}{\sum}}(u_{w}^{2}+u_{w}l_{w}+l_{w}^{2})-\frac{1}{4n^{2}}{\underset{w\in E}{\sum}}(l_{w}+u_{w})^{2}\\
&= \frac{1}{3n}{\underset{w\in E}{\sum}}(3c_{w}^{2}+r_{w}^{2}) - \frac{1}{n^{2}}{\underset{w\in E}{\sum}}c_{w}^{2}.
\end{align*}
\paragraph{Sample covariance}
\begin{align*}
Cov(X_{i},X_{j})& =\frac{1}{4n}{\underset{w\in E}{\sum}}(l_{iw}+u_{iw})(l_{jw}+u_{jw})-\frac{1}{4n^{2}}{\underset{w\in E}{\sum}}(l_{iw}+u_{iw}){\underset{w\in E}{\sum}}(l_{jw}+u_{jw})\\
&=\frac{1}{n}{\underset{w\in E}{\sum}}c_{iw}c_{jw}-\frac{1}{n^{2}}{\underset{w\in E}{\sum}}c_{iw}{\underset{w\in E}{\sum}}c_{jw}
\end{align*}
\paragraph{L2 distance}
\begin{align*}
d_{L_{2}}^{2}(x_{i},x_{j}) & ={|l_{i}-l_{j}|}^{2}+{|u_{i}-u_{j}|}^{2} \\
 &= [|(c_{i}-r_{i})-(c_{j}-r_{j})|]^{2}+[|(c_{i}+r_{i})-(c_{j}+r_{j})|]^{2} \\
 &= [(c_{i}-c_{j})-(r_{i}-r_{j})]^{2}+[(c_{i}-c_{j})+(r_{i}-r_{j})]^{2} \\
 &= (c_{i}-c_{j})^2 -2(c_{i}-c_{j})(r_{i}-r_{j})+(r_{i}-r_{j})^{2}+(c_{i}-c_{j})^{2}+2(c_{i}-c_{j})(r_{i}-r_{j})+(r_{i}-r_{j})^{2} \\
 &=2(c_{i}-c_{j})^{2}+2(r_{i}-r_{j})^{2}.
\end{align*}
\paragraph{Hausdorff distance}
\begin{align*}
d_{Hau}(x_{i},x_{j}) & =max\{|l_{i}-l_{j}|,|u_{i}-u_{j}|\} \\
 &=max \{|(c_{i}-c_{j})-(r_{ij}-r_{j})|,|(c_{i}-c_{j})+(r_{i}-r_{j})|\} \\
 &=|c_{i}-c_{j}|+|r_{i}-r_{j}|.
\end{align*}
\subsection{Accuracy of Saddlepoint approximation}

\paragraph{Normal distribution}
The density is
\begin{align*}
f(x|\mu) & =\frac{1}{\sqrt{2\pi\sigma_{0}^{2}}}e^{\frac{(x-\mu)^{2}}{2\sigma_{0}^{2}}}\\
 & =exp\{\frac{\mu}{\sigma_{0}^{2}}x-\frac{\mu^{2}}{2\sigma_{0}^{2}}\}\frac{1}{\sqrt{2\pi\sigma_{0}^{2}}}e^{-\frac{x^{2}}{2\sigma_{0}^{2}}}\\
 & =exp\{\eta x-\frac{\sigma_{0}^{2}\eta^{2}}{2}\}\frac{1}{\sqrt{2\pi\sigma_{0}^{2}}}e^{-\frac{x^{2}}{2\sigma_{0}^{2}}}
\end{align*}

Then
\begin{align*}
K(\eta) & =\frac{\sigma_{0}^{2}\eta^{2}}{2}\\
K'(\eta) & =\sigma_{0}^{2}\eta\\
K''(\eta) & =\sigma_{0}^{2}
\end{align*}

Hence,
\begin{align*}
K'(\hat{\eta}) & =x\\
\hat{\eta} & =\frac{x}{\sigma_{0}^{2}}\\
K''(\hat{\eta}) & =\sigma_{0}^{2}
\end{align*}

So the saddlepoint approximation gives
\begin{align*}
f_{T}(a;\theta) & \sim(2\pi)^{-\frac{1}{2}}(K''(\hat{\theta}))^{-\frac{1}{2}}exp\big\{(\theta-\hat{\theta})a-(K(\theta)-K(\hat{\theta}))\big\}\\
f_{T}(x;\theta) & \approx(2\pi)^{-\frac{1}{2}}(\sigma_{0}^{2})^{-\frac{1}{2}}exp\big\{(\eta-\frac{x}{\sigma_{0}^{2}})x-(\frac{\sigma_{0}^{2}\eta^{2}}{2}-\frac{x}{2\sigma_{0}^{2}})\big\}\\
 & \approx(2\pi\sigma_{0}^{2})^{-\frac{1}{2}}exp\big\{(\eta x-\frac{\sigma_{0}^{2}\eta^{2}}{2}-\frac{x^{2}}{2\sigma_{0}^{2}})\big\}.
\end{align*}

\paragraph{Exponential distribution}
The density is
\begin{align*}
f(x|\theta) & =\theta e^{-\theta x}\\
 & =exp\{\theta(-x)+log(\theta)\}
\end{align*}

Then
\begin{align*}
K(\theta) & =-log(\theta)\\
K'(\theta) & =-\frac{1}{\theta}\\
K''(\theta) & =\frac{1}{(\theta)^{2}}
\end{align*}

Hence,
\begin{align*}
K'(\hat{\theta}) & =-x\\
\hat{\theta} & =\frac{1}{x}\\
K''(\hat{\theta}) & =x^{2}
\end{align*}

So the saddlepoint approximation gives
\begin{align*}
f_{T}(a;\theta) & \sim(2\pi)^{-\frac{1}{2}}(K''(\hat{\theta}))^{-\frac{1}{2}}exp\big\{(\theta-\hat{\theta})a-(K(\theta)-K(\hat{\theta}))\big\}\\
f_{T}(x;\theta) & \approx(2\pi)^{-\frac{1}{2}}(x^{2})^{-\frac{1}{2}}exp\big\{(\theta-\frac{1}{x})(-x)-(-log\theta+log\frac{1}{x})\big\}\\
 & \approx(2\pi)^{-\frac{1}{2}}x^{-1}exp\big\{(-\theta x+1)+log\theta-log\frac{1}{x})\big\}\\
 & \approx\frac{e}{\sqrt{2\pi}}exp\big\{\theta(-x)+log\theta)\big\}.
\end{align*}

\paragraph{Poisson distribution}
The density is
\begin{align*}
f(x|\theta) & =\frac{\theta^{x}}{x!}e^{-\theta}\\
 & =exp\{log\theta \cdot x-\theta-logx!\}\\
 & =exp\{\eta x-e^{\eta}-logx!\}
\end{align*}

Then
\begin{align*}
K(\eta) & =K'(\eta)=K''(\eta)=e^{\eta}
\end{align*}

Hence,
\begin{align*}
K'(\hat{\eta}) & =x\\
\hat{\eta} & =logx\\
K''(\hat{\eta}) & =x
\end{align*}

So the saddlepoint approximation gives
\begin{align*}
f_{T}(x;\eta) & \approx(2\pi)^{-\frac{1}{2}}(x)^{-\frac{1}{2}}exp\big\{(\eta-logx)(x)-(e^{\eta}-x)\big\}\\
 & \approx(2\pi x)^{-\frac{1}{2}}exp\big\{\eta x-e^{\eta}-logxx+x)\big\}\\
 & \approx exp\big\{\eta x-e^{\eta}\big\}\sqrt{\frac{1}{2\pi x}}(\frac{e}{x})^{x}.
\end{align*}

\paragraph{Binomial distribution}
The density is
\begin{align*}
f(x|p) & =\binom{n}{x}p^{x}(1-p)^{n-x}\\
 & =exp\{\log\frac{p}{1-p}x+n\log1-p\}\binom{n}{x}\\
 & =exp\{\eta x+n\log\frac{1}{e^{\eta}+1}\}\binom{n}{x}
\end{align*}

We have
\begin{align*}
K(\eta) & =n\log(e^{\eta}+1)\\
K'(\eta) & =\frac{ne^{\eta}}{e^{\eta}+1}\\
K''(\hat{\eta}) & =\frac{ne^{\eta}}{(e^{\eta}+1)^{2}}
\end{align*}

Hence,
\begin{align*}
K'(\hat{\eta}) & =x\\
\hat{\eta} & =log\frac{x}{n-x}\\
K''(\hat{\eta}) & =\frac{(n-x)x}{n}
\end{align*}

So the saddlepoint approximation gives
\begin{align*}
f_{T}(x;\eta) & \approx(2\pi)^{-\frac{1}{2}}(\frac{(n-x)x}{n})^{-\frac{1}{2}}exp\big\{(\eta-log\frac{x}{n-x})x-(n\log(e^{\eta}+1)-n\log(\frac{n}{n-x})\big\}\\
 & \approx(2\pi\frac{(n-x)x}{n})^{-\frac{1}{2}}exp\big\{(\eta x-n\log(e^{\eta}+1)\big\}(\frac{x}{n-x})^{-x}(\frac{n}{n-x})^{n}\\
 & \approx exp\big\{(\eta x-n\log(e^{\eta}+1)\big\}(\frac{x}{n-x})^{-x}(\frac{n}{n-x})^{n}\sqrt{\frac{n}{2\pi(n-x)x}}.
\end{align*}

\paragraph{Gamma distribution}

Suppose the shape parameter $\alpha$ is known in the Gamma density,
\begin{align*}
f(x|\beta) & =\frac{\beta^{\alpha}}{\varGamma\left(\alpha\right)}x^{\alpha-1}e^{-\beta x}\\
 & =exp\{\beta(-x)+\alpha\log\beta\}\frac{x^{\alpha-1}}{\varGamma(\alpha)}
\end{align*}

Then
\begin{align*}
K(\beta) & =-\alpha log(\beta)\\
K'(\beta) & =-\frac{\alpha}{\beta}\\
K''(\beta) & =\frac{\alpha}{(\beta)^{2}}
\end{align*}

Hence,
\begin{align*}
K'(\hat{\beta}) & =-x\\
\hat{\beta} & =\frac{\alpha}{x}\\
K''(\hat{\beta}) & =\frac{x^{2}}{\alpha}
\end{align*}

So the saddlepoint approximation gives
\begin{align*}
f_{T}(x;\theta) & \approx(2\pi)^{-\frac{1}{2}}(\frac{x^{2}}{\alpha})^{-\frac{1}{2}}exp\big\{(\beta-\frac{\alpha}{x})(-x)-(-\alpha\log\beta+\alpha\log\frac{\alpha}{x})\big\}\\
 & \approx(\frac{\alpha}{2\pi x^{2}})^{\frac{1}{2}}exp\big\{-\beta x+\alpha+\alpha\log\beta+\alpha\log\frac{x}{\alpha}\big\}\\
 & \approx exp\big\{-\beta x+\alpha\log\beta\big\}\sqrt{\frac{\alpha}{2\pi x^{2}}}e^{\alpha}(\frac{x}{\alpha})^{\alpha}.
\end{align*}


\begin{thebibliography} {10}
\bibitem{1} Alvo, Mayer, and Philip Yu (2014). Statistical methods for ranking data. Springer.
\bibitem{2} Alvo, Mayer, and Philip Yu (2018). A Parametric Approach to Nonparametric Statistics. Springer International Publishing.
\bibitem{3} Alvo, Mayer, and Hang Xu (2017). "The analysis of ranking data using score functions and penalized likelihood." Austrian Journal of Statistics 46.1: 15-32.
\bibitem{4} Andreev, Andriy, Antti Kanto, and Pekka Malo (2007). "Computational examples of a new method for distribution selection in the Pearson system." Journal of applied Statistics 34.4: 487-506.
\bibitem{5} Banerjee, Trambak, et al (2021). "Nonparametric Empirical Bayes Estimation On Heterogeneous Data."
\bibitem{6} Benhaddou, Rida (2013). "Nonparametric And Empirical Bayes Estimation Methods."
\bibitem{7} Bertrand, Patrice, and Francoise Goupil (2000). "Descriptive statistics for symbolic data." Analysis of symbolic data. Springer, Berlin, Heidelberg. 106-124.
\bibitem{8} Billard, Lynne (2008). "Some analyses of interval data." Journal of computing and information technology 16.4: 225-233.
\bibitem{9} Billard, Lynne, and Edwin Diday (2003). "Symbolic data analysis: definitions and examples." Technical Report 62 pages.
\bibitem{10} Billard, Lynne, and Edwin Diday (2003). "From the statistics of data to the statistics of knowledge: symbolic data analysis." Journal of the American Statistical Association 98.462: 470-487.
\bibitem{11} Billard, Lynne, and Edwin Diday (2006). "Descriptive statistics for interval-valued observations in the presence of rules." Computational Statistics 21.2: 187-210.
\bibitem{12} Brito, Paula (2007). "Modelling and analysing interval data." Advances in data analysis. Springer, Berlin, Heidelberg, 197-208.
\bibitem{13} Casella, George (1985). "An introduction to empirical Bayes data analysis." The American Statistician 39.2: 83-87.
\bibitem{14} Casella, George (1992). "Illustrating empirical Bayes methods." Chemometrics and intelligent laboratory systems 16.2: 107-125.
\bibitem{15} Daniels, Henry E (1954). "Saddlepoint approximations in statistics." The Annals of Mathematical Statistics: 631-650.
\bibitem{16} Daniels, Henry E (1980). "Exact saddlepoint approximations." Biometrika 67.1: 59-63.
\bibitem{17} De Carvalho, Francisco de AT, Paula Brito, and Hans-Hermann Bock (2006). "Dynamic clustering for interval data based on L2 distance." Computational Statistics 21.2 : 231-250.
\bibitem{18} De Carvalho, Francisco de AT, et al (2006). "Adaptive Hausdorff distances and dynamic clustering of symbolic interval data." Pattern Recognition Letters 27.3: 167-179.
\bibitem{19} de Souza, Renata MCR, and Francisco de AT De Carvalho (2004). "Clustering of interval data based on city-block distances." Pattern Recognition Letters 25.3: 353-365.
\bibitem{20} Diday, Edwin, and J. C. Simon (1976). "Clustering analysis." Digital pattern recognition. Springer, Berlin, Heidelberg. 47-94.
\bibitem{21} Diday, Edwin (2000). "Knowledge discovery from the symbolic data and the SODAS software." PKDD 2000 workshop on Symbolic data Analysis, Lyon, 12th September.
\bibitem{22} Diday, Edwin (2008). "The state of the art in symbolic data analysis: overview and future." Symbolic Data Analysis and the SODAS Software : 3-41.
\bibitem{23} Eberly, Lynn E., and George Casella (2003). "Estimating Bayesian credible intervals." Journal of statistical planning and inference 112.1-2 : 115-132.
\bibitem{24} Efron, Bradley (2005). "Local false discovery rates."
\bibitem{25} Efron, Bradley (2011). "Tweedie's formula and selection bias." Journal of the American Statistical Association 106.496: 1602-1614.
\bibitem{26} Efron, Bradley, and Trevor Hastie (2016). Computer age statistical inference. Vol. 5. Cambridge University Press.
\bibitem{27} Gelman, Andrew, et al (2013). Bayesian data analysis. CRC press.
\bibitem{28} Gibbs, Alison L., and Francis Edward Su (2002). "On choosing and bounding probability metrics." International statistical review 70.3: 419-435.
\bibitem{29} Glen\_b (2018). (https://stats.stackexchange.com/users/805/glen-b), How does saddlepoint approximation work?, URL: https://stats.stackexchange.com/q/191492
\bibitem{30} Goutis, Constantino, and George Casella (1999). "Explaining the saddlepoint approximation." The American Statistician 53.3: 216-224.
\bibitem{31} Gu, Jiaying, and Roger Koenker. "Empirical Bayesball remixed: Empirical Bayes methods for longitudinal data." Journal of Applied Econometrics 32.3 (2017): 575-599.
\bibitem{32} Guan, Rong, Yves Lechevallier, and Huiwen Wang (2013). "Adaptive Dynamic Clustering Algorithm for Interval-valued Data based on Squared-Wasserstein Distance."
\bibitem{33} Irpino, Antonio, and Rosanna Verde (2008). "Dynamic clustering of interval data using a Wasserstein-based distance." Pattern Recognition Letters 29.11, 1648-1658.
\bibitem{34} Johns Jr, M. V (1957). "Non-parametric empirical Bayes procedures." The Annals of Mathematical Statistics ,649-669.
\bibitem{35} Johnson, Mark E (1987). Multivariate statistical simulation: A guide to selecting and generating continuous multivariate distributions. Vol. 192. John Wiley \& Sons.
\bibitem{36} Kamishima, Toshihiro (2003). "Nantonac collaborative filtering: recommendation based on order responses." Proceedings of the ninth ACM SIGKDD international conference on Knowledge discovery and data mining.
\bibitem{37} Kendall, Maurice G. (1948). "The advanced theory of statistics. Vols. 1." The advanced theory of statistics. Vols. 1. 1.Ed. 4.
\bibitem{38} Lai, Tze Leung, Yong Su, and Kevin Haoyu Sun (2014). "Dynamic empirical Bayes models and their applications to longitudinal data analysis and prediction." Statistica Sinica , 1505-1528.
\bibitem{39} Liu, Yanchi, et al (2010). "Understanding of internal clustering validation measures." 2010 IEEE international conference on data mining. IEEE.
\bibitem{40} Maritz, Johannes S (2018). Empirical Bayes methods with applications. CRC Press.
\bibitem{41} Pearson, Karl (1895). "X. Contributions to the mathematical theory of evolution.—II. Skew variation in homogeneous material." Philosophical Transactions of the Royal Society of London.(A.) 186: 343-414.
\bibitem{42} Reid, Nancy (1988). "Saddlepoint methods and statistical inference." Statistical Science , 213-227.
\bibitem{43} Robbins, Herbert (1956). "An empirical bayes approach to statistics." Proceedings of the Third Berkeley Symposium on Mathematical Statistics and Probability. Vol. 1.
\bibitem{44} Robbins, Herbert (1963). "The empirical Bayes approach to testing statistical hypotheses." Revue de l'Institut International de Statistique , 195-208.
\bibitem{45} Robbins, Herbert (1983). "Some thoughts on empirical Bayes estimation." The Annals of Statistics , 713-723.
\bibitem{46} Robbins, Herbert (1985). "Linear empirical Bayes estimation of means and variances." Proceedings of the National Academy of Sciences of the United States of America 82.6 : 1571.
\bibitem{47} Rubin, Donald B (1981). "Estimation in parallel randomized experiments." Journal of Educational Statistics 6.4: 377-401
\bibitem{48} Xu, Hang, Mayer Alvo, and Philip Yu (2018). "Angle-based models for ranking data." Computational Statistics \& Data Analysis 121 : 113-136.
\bibitem{49} Yang, Qing, Xinming An, and Wei Pan (2019). "Computing and graphing probability values of pearson distributions: a SAS/IML macro." Source code for biology and medicine 14.1: 1-6.
\bibitem{50} Zhou, Qingping, Johnson, Mark E (2018). "An approximate empirical Bayesian method for large-scale linear-Gaussian inverse problems." Inverse Problems 34.9 : 095001.
\end{thebibliography}
\end{document}